\documentclass[usenatbib]{mn2e}
\usepackage[dvipdfmx]{graphicx}
\usepackage[dvipdfmx]{color}
\usepackage{amsmath,amssymb}
\usepackage{dcolumn}
\usepackage{bm}
\usepackage{xcolor}

\newcommand{\beq}{\begin{equation}}
\newcommand{\beqa}{\begin{eqnarray}}
\newcommand{\eeq}{\end{equation}}
\newcommand{\eeqa}{\end{eqnarray}}

\newcommand{\simgt}{\lower.5ex\hbox{$\; \buildrel > \over \sim \;$}}
\newcommand{\simlt}{\lower.5ex\hbox{$\; \buildrel < \over \sim \;$}}

\newcommand{\bd}[1]{\mbox{\boldmath $#1$}}

\title[Covariance in the tSZ--WL mass scaling relation]
{Covariance in the Thermal SZ-Weak Lensing Mass Scaling Relation of Galaxy Clusters}

\author[M. Shirasaki, D.\ Nagai, E.\ T.\ Lau]
{Masato Shirasaki$^{1}$\thanks{E-mail: masato.shirasaki@nao.ac.jp},
Daisuke Nagai$^{2,3}$,
and
Erwin T.\ Lau$^{2,3}$
\\
$^{1}$National Astronomical Observatory of Japan, 
Mitaka, Tokyo 181-8588, Japan \\
$^{2}$Department of Physics, Yale University, 
New Haven, CT 06520, USA \\
$^{3}$Yale Center for Astronomy and Astrophysics, Yale University,
New Haven, CT 06520,USA \\
}
\begin{document}

\date{}

\pagerange{\pageref{firstpage}--\pageref{lastpage}} \pubyear{2015}

\maketitle

\label{firstpage}

\begin{abstract}

The thermal Sunyaev-Zel'dovich (tSZ) effect signal is widely recognized 
as a robust mass proxy of galaxy clusters with small intrinsic scatter. 
However, recent observational calibration 
of the tSZ scaling relation using weak lensing (WL) mass 
exhibits considerably larger scatter than the intrinsic scatter 
predicted from numerical simulations. 
This raises a question as to whether we can realize
the full statistical power of ongoing and upcoming 
tSZ-WL observations of galaxy clusters. 
In this work, we investigate the origin of observed scatter in the tSZ-WL scaling relation, 
using mock maps of galaxy clusters extracted from cosmological hydrodynamic simulations. 
We show that the inferred intrinsic scatter from mock tSZ-WL analyses is considerably 
larger than the intrinsic scatter measured in simulations, and comparable to the scatter 
in the observed tSZ-WL relation. We show that this enhanced scatter originates from the 
combination of the projection of correlated structures along 
the line of sight and the uncertainty in the cluster radius associated 
with WL mass estimates, causing the amplitude of the scatter to depend on 
the {\em covariance} between tSZ and WL signals. We present a statistical model to 
recover the unbiased cluster scaling relation and cosmological parameter by taking 
into account the covariance in the tSZ-WL mass relation from multi-wavelength cluster surveys. 
\end{abstract}


\section{INTRODUCTION}

In recent years, the Sunyaev-Zel'dovich (SZ) effect observations of galaxy clusters have emerged as a powerful probe of the growth of cosmic structure and cosmology.  The thermal SZ (tSZ) effect is the inverse Compton scattering of the CMB photons off of energetic electrons in the intracluster medium (ICM) \citep{1972CoASP...4..173S}. Since the SZ effect signal is independent of redshift, it offers a powerful way of detecting galaxy clusters out to high redshift with the current generation of microwave experiments, such as the Atacama Cosmology Telescope (ACT), the South Pole Telescope (SPT), and the Planck satellite \citep[e.g.,][]{2013JCAP...07..008H,2015ApJS..216...27B,2015arXiv150201598P}. These cluster samples have been used to measure the evolution of cluster abundance over the cosmic time and constrain cosmological parameters \citep[e.g.,][]{2013JCAP...10..060S,2015arXiv150201597P,2016arXiv160306522D}. 

Cosmological constraints derived from these surveys rely critically on the calibration of the relationship between the observable and mass of galaxy clusters. Numerical simulations predict that the tSZ effect signal is a robust proxy of cluster mass with intrinsic scatter of $\lesssim 10\%$ as it directly probes the thermal energy content of the virialized ICM \citep[e.g.,][]{2005ApJ...623L..63M, 2006ApJ...650..538N, 2012MNRAS.422.1999K,2013MNRAS.429..323S,2015ApJ...807...12Y}.

However, the cluster-based cosmological constraint hinges on the still poorly understood calibration of the relationship between the observable and cluster mass \citep[e.g.,][]{2015ApJ...799..214B,2015arXiv151200910S}.  As such, the tSZ-mass scaling relation has been calibrated observationally, based on the assumption that the cluster gas is in hydrostatic equilibrium with the gravitational potential of galaxy clusters. However, the hydrostatic mass estimate derived from X-ray observations is shown to produce biased estimates of cluster mass at the level of $5-30\%$ depending on their dynamical states \citep[e.g.,][]{2006MNRAS.369.2013R,2007ApJ...655...98N}, and it is one of the dominant sources of astrophysical uncertainties in cosmological constraints from SZ surveys \citep[e.g.,][]{2014A&A...571A..20P,2015arXiv150201597P}.

Weak lensing (WL) mass measurements, which directly probe the projected mass distribution of the cluster, provide a promising way to measure cluster mass independently of their dynamical states \citep[e.g., ][]{2009ApJ...701L.114M, 2009MNRAS.399L..84M, 2012ApJ...758...68H, 2012MNRAS.427.1298H, 2013MNRAS.429.3627M, 2014MNRAS.443.1973V, 2014ApJ...785...20J, 2014MNRAS.442.1507G, 2015arXiv150908930B, 2015arXiv151101919S}. However, recent tSZ and WL measurements suggest that the scatter in the tSZ-WL mass scaling relation is on the order of $\sim20\%$ \citep[e.g.,][]{2012ApJ...754..119M},  which is considerably larger than the intrinsic scatter predicted by numerical simulations. This raises a question as to whether the WL mass calibration of the SZ-selected clusters can realize the full statistical power of the ongoing and upcoming SZ surveys to test cosmological models. 

In this work, we investigate the origin of the large discrepancy between the intrinsic scatters in the tSZ-mass scaling relation from simulations and observations, by using mock tSZ and WL analyses of galaxy clusters extracted from high-resolution cosmological hydrodynamic simulation. We show that most of the scatter in the observed tSZ-WL mass relation is driven by the combination of the enhanced scatter in tSZ due to projections of correlated structures in the outskirt of individual clusters and the bias in WL determined cluster radius, within which the tSZ signal is measured. Most importantly, our results demonstrate the importance of the covariance between tSZ and WL due to the correlated structures along 
the line of sight.  
We present a statistical model to recover the unbiased $Y-M$ relation from a set of tSZ and WL measurements, by taking into account covariances among clusters' observables.

The paper is organized as follows. In Section~\ref{sec:sim}, we describe our simulations and mock tSZ and WL analyses 
of simulated clusters. We first examine the nature of scatters in tSZ and WL measurements in Section~\ref{sec:scatter} and 
the covariance between tSZ and WL observables and its impact on cluster-based cosmological analyses 
in Section~\ref{sec:covariance}. Section~\ref{sec:baryon} explores the systematic uncertainties associated with baryonic effects.
Conclusions are summarized in Section~\ref{sec:conclusions}.

\section{METHODS}\label{sec:sim}

\subsection{Hydrodynamic Simulations}

\begin{figure}
\centering \includegraphics[clip, width=0.95\columnwidth]{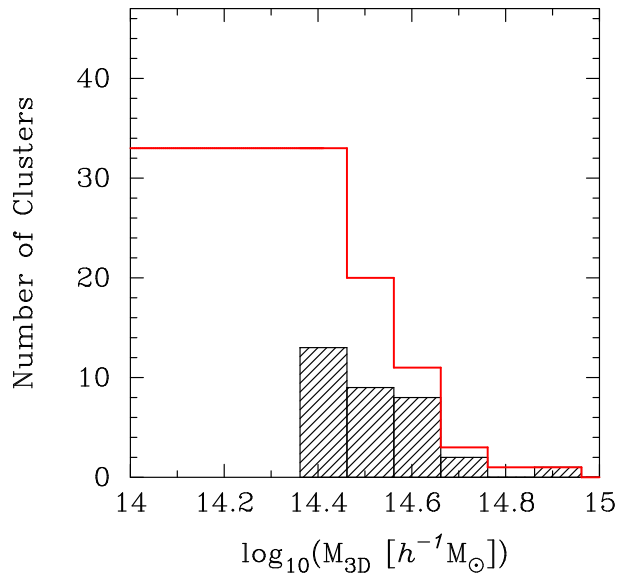}
\caption{
	The distribution of halo mass for 
	our simulated clusters at $z=0.33$, where 
	the halo mass is defined by the enclosed mass within 
	the radius at which the mean interior density 
	equals 500 times the critical density of the universe.
	The black hatched histogram represents the differential distribution,
	while the red histogram shows the cumulative distribution.
	Note that we show the number of clusters with a bin size 
	of $\Delta \log M_{\rm 3D} = 0.1$ in this figure.
	\label{fig:dist_mass_3d}
	}
\end{figure}

In this work, we analyze the mass-limited sample of 33 galaxy clusters extracted from 
the {\it Omega500} non-radiative (NR) hydrodynamics \citep{2014ApJ...782..107N} in a 
flat $\Lambda$CDM model with the WMAP five-year results \citep{2009ApJS..180..330K}:
$\Omega_{\rm m0} = 0.27$ (matter density), $\Omega_{\rm b0} = 0.0469$ (baryon density),
$H_{0} = 100h =70\, {\rm km}\, {\rm s}^{-1}{\rm Mpc}^{-1}$ (Hubble constant),
and $\sigma_{8} = 0.82$ (the mass variance within a sphere with a radius of 8 $h^{-1}\, {\rm Mpc}$).
The simulation is performed using the Adaptive Refinement Tree (ART)
$N$-body+gas-dynamics code 
\citep{1999PhDT........25K, 2002ApJ...571..563K, 2008ApJ...672...19R}, 
which is an Eulerian code that uses adaptive refinement in space and time 
and non-adaptive refinement in mass \citep{2001ApJ...554..903K} to achieve 
the dynamic range necessary to resolve the cores of halos formed in self-consistent 
cosmological simulations.  The simulation volume has a comoving box length of 500~$h^{-1}\, {\rm Mpc}$,
resolved using a uniform $512^3$ root grid and 8 levels of mesh refinement, 
implying a maximum comoving spatial resolution of 3.8~$h^{-1}\, {\rm kpc}$.
While the effects of baryonic physics, such as radiative gas cooling, star formation and 
energy feedback from supernovae and active galactic nuclei are important in the cluster 
core regions ($r \simlt 0.15 R_{\rm 500c}$), these additional physics are shown to have 
negligible ($\lesssim 2$\%) impact on the scatter in the tSZ-mass scaling relation 
\citep{2006ApJ...650..538N,2012ApJ...758...74B,2012MNRAS.422.1999K}. 
In Section~\ref{sec:baryon}, we assess the impact of baryonic physics with 
the {\em Omega500} simulation that includes radiative cooling, star formation, and supernova feedback. 

Cluster-sized halos are identified in the simulation using a spherical overdensity halo 
finder described in \citet{2014ApJ...782..107N}. 
We define the three-dimensional (3D) mass of cluster using the spherical overdensity 
criterion: $M_{\rm 500c} = 500 \rho_{\rm crit}(z) (4\pi/3)R_{\rm 500c}^3$,  
where $\rho_{\rm crit}(z)$ is the critical density of the universe at a given redshift $z$.
In the following, we denote this 3D mass as $M_{\rm 3D}$. We select clusters with 
$M_{\rm 3D}\geq 3 \times 10^{14}\, h^{-1}M_{\odot}$ at 
$z=0$ and re-simulate the box with higher resolution dark matter particles 
in regions of the selected clusters with the ``zoom-in'' technique, resulting in an effective
mass resolution of $2048^{3}$, corresponding to 
a dark matter particle mass of  $1.09\times10^{9}\, h^{-1} M_{\odot}$, 
inside spherical region with cluster-centric radius of 
three time the virial radius for each cluster. 

In this work, we work mainly with a mass-limited sample of $33$ clusters with 
$M_{\rm 500c}\geq 2.3\times10^{14}\, h^{-1} M_{\odot}$ at $z=0.33$, 
which is comparable to the typical redshift of recent WL cluster observations 
\citep[e.g.,][]{2012ApJ...758...68H,2015arXiv150908930B}.
Figure \ref{fig:dist_mass_3d} shows the mass distribution of our selected clusters at $z=0.33$.

\subsection{Mock Maps}
In this section we describe our procedure 
for creating mock lensing and tSZ maps 
from cosmological hydrodynamic simulations.

\subsubsection{Weak lensing maps}
In gravitational lensing, the distortion of image of a source 
object with true angular position $\mbox{\boldmath $\beta$}$ and 
observed angular position $\mbox{\boldmath $\theta$}$ can be
characterized by the following $2\times2$ matrix:
\beqa
A_{ij} = \frac{\partial \beta^{i}}{\partial \theta^{j}}
           \equiv \left(
\begin{array}{cc}
1-\kappa -\gamma_{1} & -\gamma_{2}  \\
-\gamma_{2} & 1-\kappa+\gamma_{1} \\
\end{array}
\right), \label{distortion_tensor}
\eeqa
where $\kappa$ is convergence and $\gamma$ is shear.

One can relate each component of $A_{ij}$ to
the second derivative of the gravitational potential $\Phi$ of the lens object as follows
\citep{Bartelmann2001, Munshi2008};
\beqa
A_{ij} &=& \delta_{ij} - \phi_{ij}, \label{eq:Aij} \\
\phi_{ij}  &=&\frac{2}{c^2}\int _{0}^{\chi}{\rm d}\chi^{\prime} g(\chi,\chi^{\prime}) \partial_{i}\partial_{j}\Phi(\chi^{\prime}), \label{eq:shear_ten}\\	
g(\chi,\chi^{\prime}) &=& \frac{r(\chi-\chi^{\prime})r(\chi^{\prime})}{r(\chi)},
\eeqa
where $\chi$ is the comoving distance and $r(\chi)$ is the comoving angular diameter distance.
Gravitational potential $\Phi$ can then be related to the matter density perturbation $\delta$ by the Poisson equation.

The convergence can then be expressed as the weighted integral of $\delta$ along the line of sight,
\beqa
\kappa = \frac{3}{2}\left(\frac{H_{0}}{c}\right)^2 \Omega_{\rm m0} \int _{0}^{\chi}{\rm d}\chi^{\prime} g(\chi,\chi^{\prime}) \frac{\delta}{a}. \label{eq:kappa_delta}
\eeqa
The relation between convergence and shear in Fourier space is given by 
\beqa
\tilde{\gamma}(\mbox{\boldmath $k$}) &=& \tilde{\gamma_{1}}(\mbox{\boldmath $k$})+i\tilde{\gamma_{2}}(\mbox{\boldmath $k$}) 
= \frac{k_{1}^{2}-k_{2}^2+ik_{1}k_{2}}{k^2} \tilde{\kappa}(\mbox{\boldmath $k$}), \label{eq:shear-kappa_ft} \\
\tilde{\kappa}(\bd{k}) &=& \tilde{\gamma_1}(\bd{k})\cos 2\phi_{\bd{k}} + \tilde{\gamma_2}(\bd{k})\sin 2\phi_{\bd{k}}, \label{eq:shear-kappa_ft2}
\eeqa
where $\tilde{X}(\mbox{\boldmath $k$})$ is the Fourier coefficient of $X(\mbox{\boldmath $\theta$})$ 
and $\bd{k} = (k_1,k_2) = k(\cos \phi_{\bd{k}}, \sin \phi_{\bd{k}})$.

To simulate WL cluster mass measurement, 
we first create projected mass density maps of each cluster viewed along 
three orthogonal projections, $(x,y,z)$. 
We then derive the convergence field using Eq.~(\ref{eq:kappa_delta})
and transform convergence into shear using Eq.~(\ref{eq:shear-kappa_ft}).
Throughout this paper, we consider a single source redshift $z_s = 1$ for
lensing calculations.
We then generate the projected mass density map on the $2048^{2}$ 
two-dimensional mesh points by extracting all particles around each cluster 
in a comoving box with volume of
$15.6 \times 15.6 \times L_{\rm depth} \, (h^{-1}{\rm Mpc})^3$,
where $L_{\rm depth}$ is the projection depth along the line of sight. 
We vary the projection depth
$L_{\rm depth}=10, 20, 100$, and $500\, h^{-1}{\rm Mpc}$
to explore the effects of correlated structures along the line of sight, 
while keeping the transverse size of the analysis volume fixed.
Note that we ignore two ``past lightcone" effects associated with 
(1) the evolution of large-scale structure and (2) the increasing transverse 
size with redshift, which requires ray-tracing simulation \citep[e.g.,][]{White2000} 
and left for future work.

Because the dark matter particles come in different masses in our 
zoom-in simulations, the mass density maps with a large projection 
depth get contribution from low resolution dark matter particles,
which appear as localized, point like masses. 
We alleviate this effect by smoothing the mass associated with these particles 
uniformly over the mesh in which these particles reside. 
We confirmed that the average WL signal in the radial range of $0.5<R/R_{\rm 500c}<2$ 
converges to better than $1$\% in the four different cases of $L_{\rm depth}$.
We, therefore, conclude that the low resolution dark matter particles 
do not affect the resulting mean value of the WL-inferred mass in 
the radial range of our interest.

\begin{figure}
\centering \includegraphics[clip, width=0.85\columnwidth]
{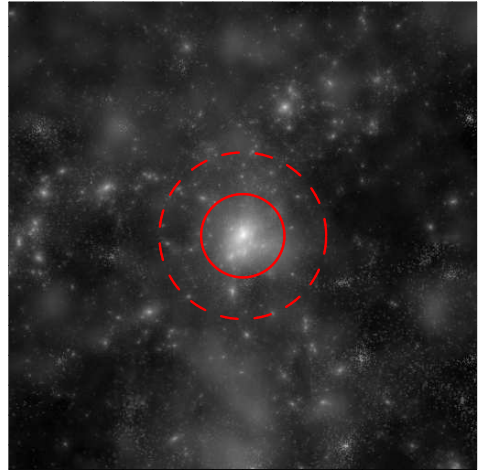} \\
\centering \includegraphics[clip, width=0.74\columnwidth]
{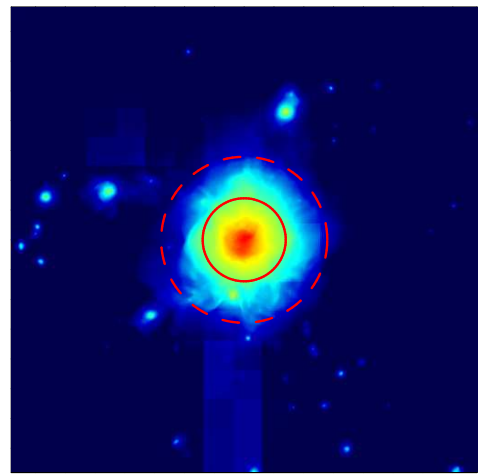}
\caption{
	Projected total mass density ({\em top} panel)
	and in Compton-$y$ ({\em bottom} panel) maps
	of the simulated cluster with	
	$M_{\rm 500c}=9.1\times10^{14}$\, $h^{-1} M_{\odot}$.
	The projection depth is set to $500 h^{-1} {\rm Mpc}$.
	In each panel, the solid and dash circles indicate 
	the radius of $R_{\rm 500c}$ and $2R_{\rm 500c}$,
	respectively. 
	\label{fig:image_map}
	}
\end{figure} 

\subsubsection{Compton-$y$ maps}

The tSZ effect is a spectral distortion of CMB caused by inverse Compton scattering 
of CMB photons off of electrons in the high-temperature plasma in the ICM. 
The temperature change at frequency $\nu$ of the CMB is given by 
$\Delta T_{\nu}/T_{\rm CMB}=f_{\nu}(x) y$, where 
$f_{\nu}(x)=[x(e^x+1)/(e^x-1)-4](1+\delta_{\rm SZE}(x,T_e))$ is a frequency dependent factor, 
$\delta_{\rm SZE}(x,T_e)$ is the frequency dependent relativistic correction 
and $x \equiv h\nu/k_BT_{\rm CMB}$. The amplitude of 
the SZE signal is given by the Compton-$y$ parameter:
\beqa
y \equiv \frac{\sigma_{\rm T}}{m_{e}c^2} \int {\rm d}\ell\, P_{e}(\ell), 
\label{eq:def_y}
\eeqa
where $\sigma_T$ is the Thomson cross-section, $m_e$ is the electron rest mass, $c$ 
is the speed of light, $P_e$ is the electron pressure, and the integral is performed along 
the line of sight $\ell$.

We generate the tSZ maps of each cluster viewed along three orthogonal projections 
on $2048^2$ mesh points by integrating Eq.~(\ref{eq:def_y}) in a comoving box with 
volume of $15.6 \times 15.6 \times L_{\rm depth} \, (h^{-1}{\rm Mpc})^3$, 
where $L_{\rm depth}=10, 20, 100$, and $500\, h^{-1}{\rm Mpc}$.
Because of the AMR nature of the simulation, the gas in the low density region
is not refined as aggressively and appear as a grid-like feature in the map.
However, we checked that most grid-like features are found in the outer region 
of clusters ($R \gtrsim 2R_{\rm500c}$) and hence do not affect our analyses. 
An example of the resulting Compton-$y$ map is shown in Figure \ref{fig:image_map}.

\subsection{Profile Fitting}
\label{subsec:fitting}

\begin{table}
\begin{tabular}{@{}lccccc|}
\hline
\hline
& Model & Free parameters & Fitting range \\ \hline
WL & NFW (Eq.~\ref{eq:NFW_profile}) & $M_{\rm 2D}$ and $c_{{\rm 500c},h}$ & $0.5'-10'$ \\
tSZ & gNFW (Eq.~\ref{eq:UPP}) & $M_{500,p}$ and $c_{500,p}$ & $0.1'-5'$ \\
\hline
\end{tabular}
\caption{
	Summary of our $\chi^2$ fitting for WL and tSZ measurement
	of clusters. 
	}
\label{tab:fitting}
\end{table}

From both WL and tSZ maps, we measure the azimuthally averaged, logarithmically spaced
radial profiles in the radial range of $\theta = 0.1'-30'$ around the center of each cluster. 
Note that the angular size of $R_{\rm 500c}$ corresponds to $2'-3'$ for clusters at $z=0.33$, 
which is well within the range of our angular bins.

In order to find the best representation of an observable $X(\theta)$,  
where $X$ can be our WL shear or Compton-$y$, we use a $\chi^{2}$-fitting metric and the non-linear 
least-squares Levenberg-Marquardt algorithm \citep{1992nrfa.book.....P}.
Suppose that the expected signal is expressed as 
$X_{\rm model}(\theta; {\bd p})$ with a set of parameters ${\bd p}$, 
a $\chi^{2}$ metric can then be defined as 
\beqa
\chi^{2} = \sum_{i=1}^{N_{\rm bin}}
(X(\theta_{i})-X_{\rm model}(\theta_{i}; {\bd p}))^2,
\eeqa
where $N_{\rm bin}$ is the number of angular bins.

For WL maps, the observable is the tangential component of reduced shear around each cluster, defined as
\beqa
g_{T} = -\frac{\gamma_1}{1-\kappa}\cos 2\phi - 
\frac{\gamma_2}{1-\kappa}\sin 2\phi,
\eeqa
where $\phi$ is the azimuthal angle on each WL map.
To model $g_{T}$, we use the NFW profile 
for matter density profile, which is
given by
\beqa
\rho_{h}(r) = \frac{\rho_{s}}{(r/r_{s})(1+r/r_s)^2},
\label{eq:NFW_profile}
\eeqa
where $\rho_s$ and $r_s$ are the scale density and the scale radius, respectively, and the concentration 
parameter is defined as $c_{{\rm 500c},h}\equiv R_{\rm 500c}/r_{s}$ \citep{Navarro1997}.
The corresponding convergence and shear can then be obtained analytically \citep{2000ApJ...534...34W}.
We denote $M_{\rm 2D}$ as $M_{\rm 500c}$ inferred from $\chi^2$ fitting to $g_{T}$. 
When performing $\chi^2$ fitting, we consider the angular range of $0.5'-10'$,
because it is difficult to simulate gravitational lensing effect with our method in the inner 
region of $\theta \lesssim 0.5'$, while the correlated matter contribution dominates at $\theta \gtrsim 10'$.

\begin{figure}
\centering \includegraphics[clip, width=0.85\columnwidth]
{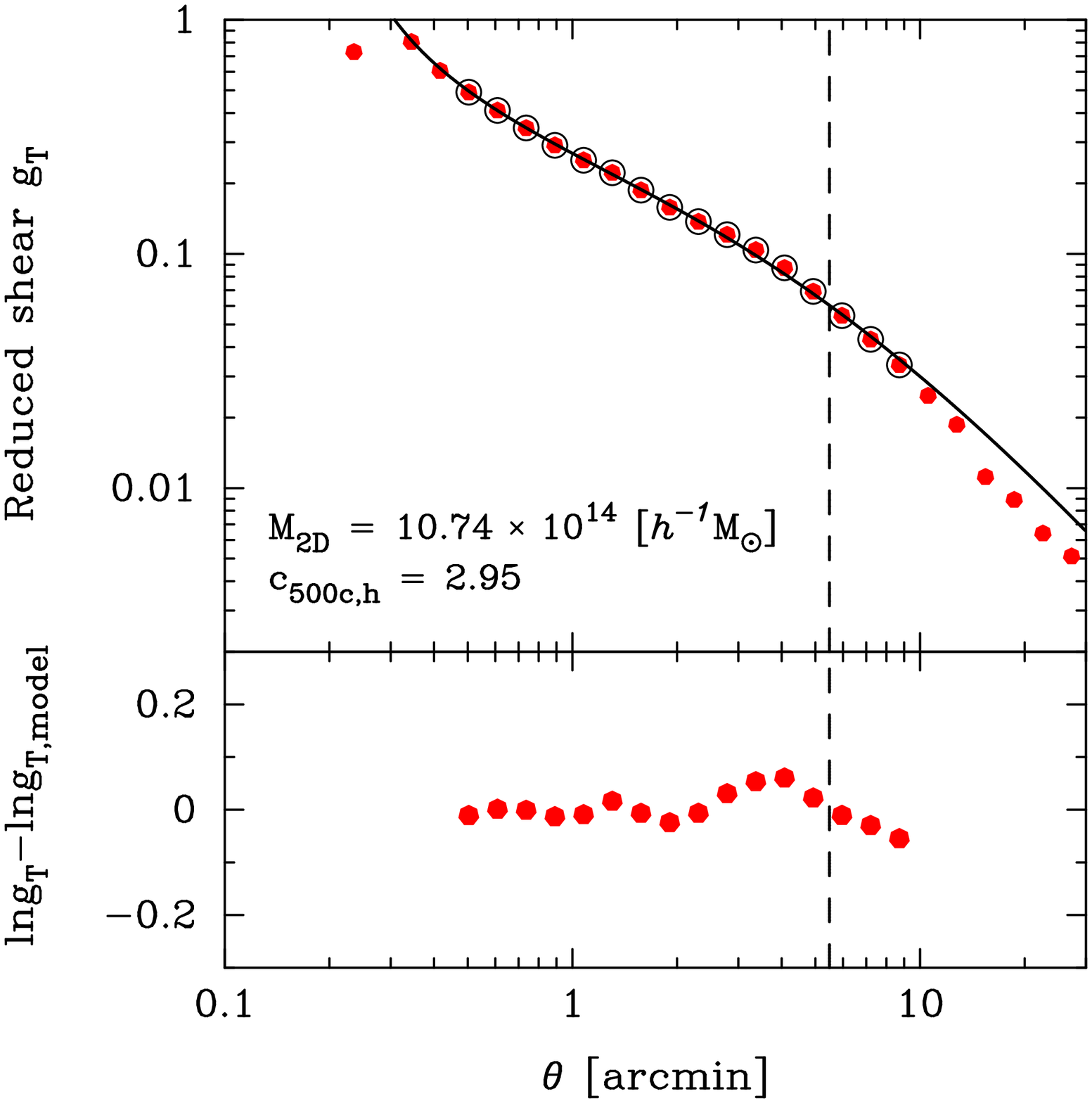} \\
\centering \includegraphics[clip, width=0.85\columnwidth]
{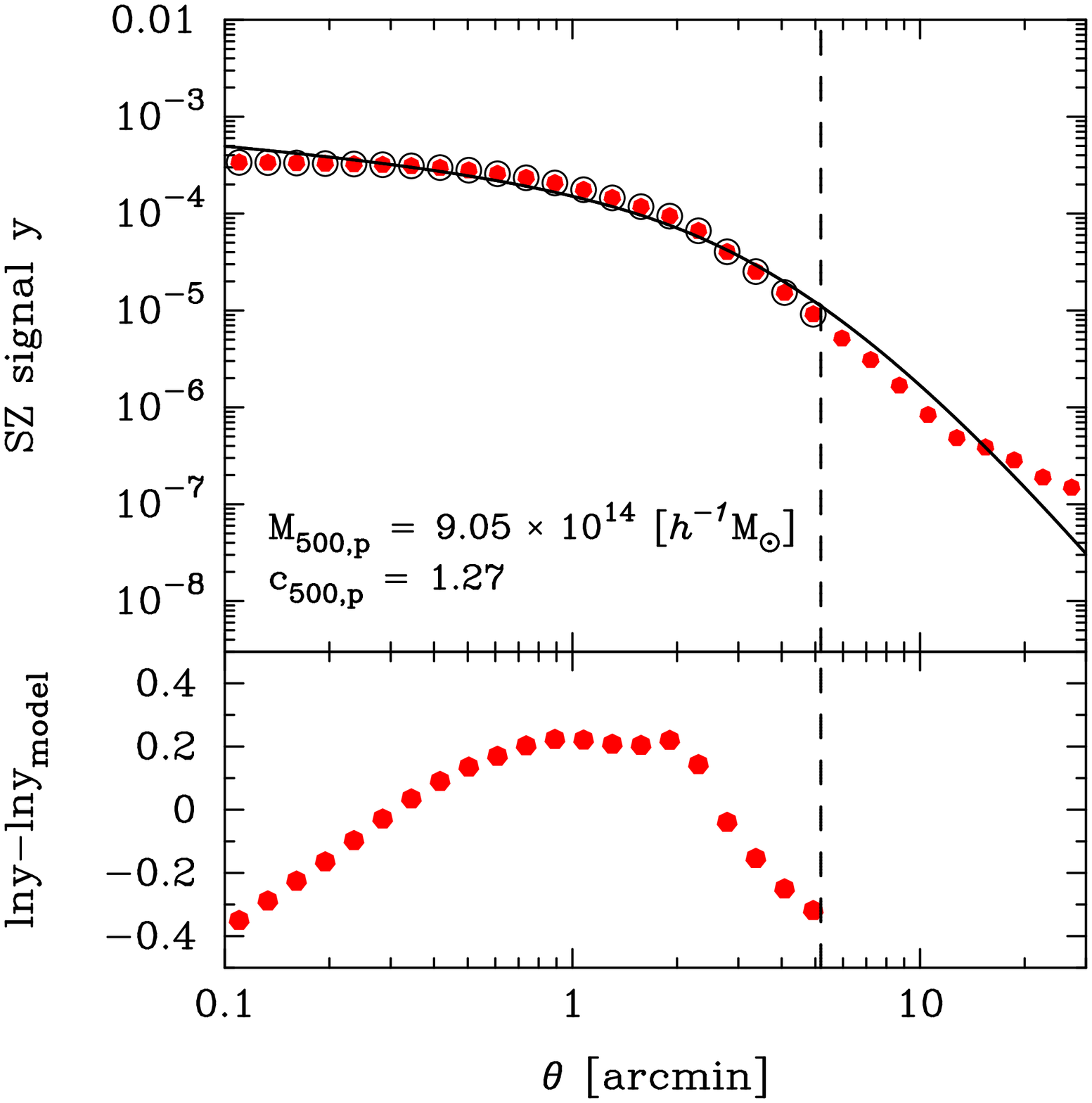}
\caption{
	An example of profile fitting to 
	reduced shear ({\em top} panel)
	and Compton-$y$ ({\em bottom} panel) profile.
	We work on the simulated cluster with
	$M_{\rm 500c}=9.1\times10^{14}$\, $h^{-1} M_{\odot}$
	(as same as shown in Figure~\ref{fig:image_map}).
	The projection depth is set to $500 h^{-1} {\rm Mpc}$.
	In each panel, the red points represent the measured profile
	and black open circles show the bins used for profile fitting.
	The solid line in upper portion is the best-fitted profile and
	the residual is shown in bottom portion of each panel. 
	The dash line in each panel corresponds to the radius 
	obtained from $M_{\rm 2D}$ or $M_{500,p}$. 
	The best-fitted value of parameters is summarized in each panel.
	\label{fig:fit_prof_example}
	}
\end{figure} 

The observable in tSZ maps is the azimuthally averaged Compton-$y$ profile around each cluster.
To model this profile, we use the generalized NFW (gNFW) pressure profile \citep{2007ApJ...668....1N}.
Since our ultimate goal is to apply the method developed in this paper to real cluster observations, we adopt the 
universal pressure profile calibrated by X-ray observations of nearby clusters \citep{2010A&A...517A..92A},
which is given by
\beqa
&P_{e}(r) = 1.88\times E(z)^{8/3} 
\left(\frac{M_{500,p}}{10^{14}h^{-1}M_{\odot}}\right)^{0.787+\alpha^{\prime}_{\rm P}(r/R_{500,p})} \nonumber \\ 
&\times h^2 \,p(r/R_{500,p})\, {\rm eV}\, {\rm cm}^{-3}, \label{eq:UPP}
\eeqa
where $E(z) = (\Omega_{\rm m0}(1+z)^3 + 1-\Omega_{\rm m0})^{1/2}$ and $R_{500,p}$ is defined by the relation of 
$M_{500,p} = 4\pi R_{500,p}^3 \times 500\rho_{\rm crit}(z)/3$.
In Eq.~(\ref{eq:UPP}), the functional form of $p(x)$ and 
$\alpha^{\prime}_{\rm P}(x)$ are specified by
\beqa
p(x) &=& \frac{P_{0}h^{-3/2}}{(c_{500,p}x)^{\gamma}(1+[c_{500,p}x]^{\alpha})^{(\beta-\gamma)/\alpha}}, \\ \label{eq:gNFW}
\alpha^{\prime}_{\rm P}(x) 
&=& 0.10-0.22\left[\frac{\left(x/0.5\right)^{3}}{1+\left(x/0.5\right)^{3}}\right].
\eeqa
Throughout our analysis, 
we use the best-fit parameters derived from 
all the {\tt REXCESS} data set in \citet{2010A&A...517A..92A}:
$P_{0}=4.921$, $\gamma=0.3081$, $\alpha=1.0510$ 
and $\beta=5.4905$ but float the parameters $M_{500,p}$ and $c_{500,p}$.
A $\chi^{2}$-fitting with Eq.~(\ref{eq:UPP}) is performed in the angular range of $0.1'-5'$.\footnote{
Note that our results slightly depend on the fitting range. 
The fractional change in the scatter in $Y_{\rm 2D}$ 
is of order 5\% when using the fitting range of $0.1'-7'$.}
Note that our results are insensitive to the choice of the assumed pressure 
profile; e.g., the fractional change in the scatter in $Y$ is less than $2\%$ 
if we use the pressure profile calibrated based on the 
NR simulations \citep{2007ApJ...668....1N}.
Table~\ref{tab:fitting} summarizes 
the parameters of the $\chi^2$-fitting to 
the mock WL and tSZ maps of the simulated clusters. 
Figure~\ref{fig:fit_prof_example} shows an example 
of profile fitting results to our sample.

\section{Scatters in tSZ and WL measurements}
\label{sec:scatter}

\subsection{3D $Y-M$ Relation}
\label{subsec:prop_Y3D}

First, we quantify the {\em intrinsic} scatter in the tSZ-WL mass scaling relation, using 
the spherically integrated global tSZ signal and true cluster mass computed 
directly from the simulation. The global tSZ signal is represented by 
the integrated Compton-$y$ parameter $Y_{\rm 3D}$, which is the volume 
integrated electron pressure in the ICM within a sphere with a radius $R_{\rm ref}$:
\beqa
Y \equiv \frac{\sigma_{\rm T}}{m_{e}c^2}
\int_{0}^{R_{\rm ref}}\, P_{e}(r)4\pi r^2{\rm d}r,
\label{eq:def_Y}
\eeqa
where $R_{\rm ref}$ is a reference radius to define the boundary of clusters. 
We evaluate $Y$ using the spherically averaged electron pressure profile $P_{e}$ 
of each simulated cluster, and we set $R_{\rm ref} =R_{\rm 500c}$ which is obtained 
from the true 3D mass, $M_{\rm 3D}$, computed directly from simulation. 
Hereafter, we denote to this spherically averaged $Y$ as $Y_{\rm 3D}$.

\begin{figure}
\centering \includegraphics[clip, width=1.0\columnwidth]{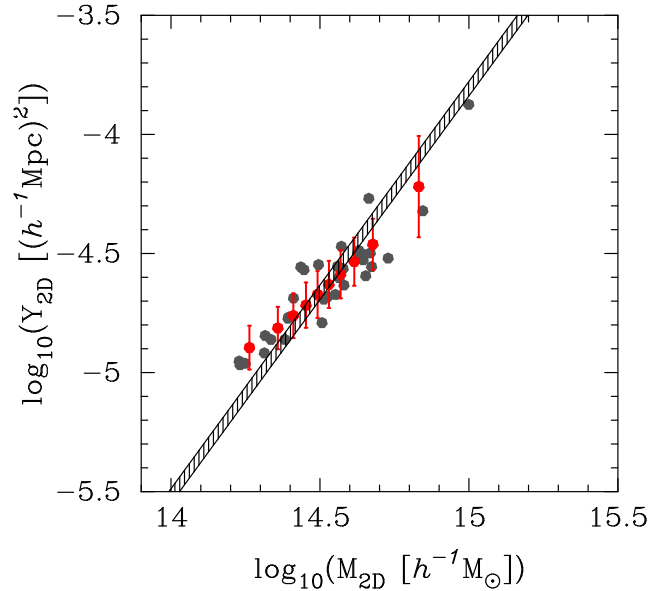}
\caption{
	The scatter plot of the $Y-M$ relation inferred 
	from two-dimensional tSZ and WL mock observations.
	The gray points represent the result of $Y_{\rm 2D}$ 
	and $M_{\rm 2D}$ obtained from a $\chi^2$ fitting 
	over 33 realizations of tSZ and WL maps.
	The black hatched region corresponds to 
	the underlying $Y-M$ relation 
	for our simulated clusters with 
	the scatter of $\sigma_{\log Y} = 0.030$,
	or $\sigma_{\ln Y} = \sigma_{\log Y} \times \ln 10 = 6.9\%$. 
	The red point with error bar shows our modeling of the 
	$Y_{\rm 2D}-M_{\rm 2D}$ relation.
	\label{fig:scat_M2D_Y2D}
	}
\end{figure}

Performing a linear least square fitting to $33$ clusters in our sample at $z=0.33$, 
the best-fit scaling relation between $\log Y_{\rm 3D}$ and $\log M_{\rm 3D}$ is
\beqa
\log \left(\frac{Y_{\rm 3D}}{(h^{-1}\,{\rm Mpc})^2}\right)= 
1.71\log \left(\frac{M_{\rm 3D}}{10^{14}\,h^{-1}\,M_{\odot}}\right)
-5.51,
\label{eq:true_Y-M}
\eeqa
where the $1\sigma$ errors in the normalization and slope 
are found to be 0.013 and 0.025, respectively.
Hence, the best-fit slope is consistent with the self-similar 
prediction of $5/3$ within $2\sigma$.
The best-fit relation is shown as hatched region 
in Figure~\ref{fig:scat_M2D_Y2D}.  

We quantify the intrinsic scatter\footnote{Throughout the paper, we use $\log = \log_{10}$ to compute 
scatter in scaling relations unless noted otherwise.} of the 
$Y_{\rm 3D}-M_{\rm 3D}$ relation as 
\beqa
\sigma_{\log  Y, {\rm 3D}}^2 = \frac{1}{N_{{\rm s}}-1}  \sum_{i=1}^{N_{\rm s}} 
\left( \log Y_{{\rm 3D},i}-\log Y_{{\rm 3D, fit}}(M_{{\rm 3D},i}) \right)^2,
\label{eq:scat_Y3D}
\eeqa
where $N_{\rm s}=33$ and $Y_{{\rm 3D, fit}}$ denotes the best-fit relation given by Eq.~(\ref{eq:true_Y-M}).
The intrinsic scatter is $\sigma_{\log Y, {\rm 3D}}=0.030$ or 
$\sigma_{\ln Y, {\rm 3D}}= \sigma_{\log Y, {\rm 3D}}\times \ln10= 6.9\%$ 
for our sample at $z=0.33$, and it is consistent with previous results based on cosmological hydrodynamic simulations
\citep[e.g.,][]{2006ApJ...650..538N, 2010ApJ...725.1124Y, 2010ApJ...715.1508S, 2012MNRAS.419.1766K, 2012ApJ...758...74B,2015ApJ...807...12Y}.

\subsection{2D $Y-M$ relation from tSZ and WL maps}
\label{subsec:prop_Y2D}

Next, we consider the $Y-M$ scaling relation measured from the projected tSZ and WL mass maps. 
Following the procedures described in Section~\ref{subsec:fitting}, 
we fit the Compton-$y$ profile of each simulated cluster using
the projected gNFW profile (see Eq.~\ref{eq:UPP})
to obtain the parameters of $M_{500,p}$ and $c_{500,p}$.
We then compute the integrated Compton-$y$ parameter $Y$ using Eq.~(\ref{eq:def_Y}) with the fitted result of $M_{500,p}$ and $c_{500,p}$ as the parameters of $P_{e}(r)$
and the radius 
$R_{\rm ref}=R_{\rm 2D}$ inferred from the WL mass $M_{\rm 2D}$ as the outer boundary of the cluster. 
We denote this $Y$ measurement as $Y_{\rm 2D}$ and use the projection depth which is 
matched to the size of the entire simulation box $L_{\rm depth}=500\, h^{-1}\,{\rm Mpc}$ 
for both $Y_{\rm 2D}$ and $M_{\rm 2D}$ measurements. 

We derive the values of $Y_{\rm 2D}$ and $M_{\rm 2D}$ over $33$ realizations of WL and 
tSZ maps and compare them with the true $Y_{\rm 3D}-M_{\rm 3D}$ scaling relation 
from Eq.~(\ref{eq:true_Y-M}). Figure~\ref{fig:scat_M2D_Y2D} shows that the $Y_{\rm 2D}-M_{\rm 2D}$ 
scaling relation (indicated by gray points) exhibits considerably larger scatter than the underlying 
$Y_{\rm 3D}-M_{\rm 3D}$ relation (indicated by the hatched region).
The scatter in the $Y_{\rm 2D}-M_{\rm 2D}$ relation is $\sigma_{\log Y, {\rm 2D}} = 0.10$,
which is larger than the 3D case by a factor of $3$. 
This level of scatter is consistent with observations \citep[e.g., ][]{2009ApJ...701L.114M, 2009MNRAS.399L..84M, 2012ApJ...758...68H, 2012ApJ...754..119M, 2012MNRAS.427.1298H, 2014MNRAS.442.1507G}.

Furthermore, we perform a least square fitting to 33 clusters in order to
find the best-fit relation between $\log {Y_{\rm 2D}}$ and 
$\log M_{\rm 2D}$. For the projection depth of 500 $h^{-1}\, {\rm Mpc}$,
the best-fit normalization and slope for the $x$-axis projection 
are found to be $-5.24\pm0.050$ and $1.17\pm0.090$ 
($1\sigma$ error), respectively. Note that the best-fit normalization and slopes are consistent 
among three orthogonal projections at 1$\sigma$ level.

\begin{figure*}
\centering 
\includegraphics[clip, width=1.0\columnwidth]{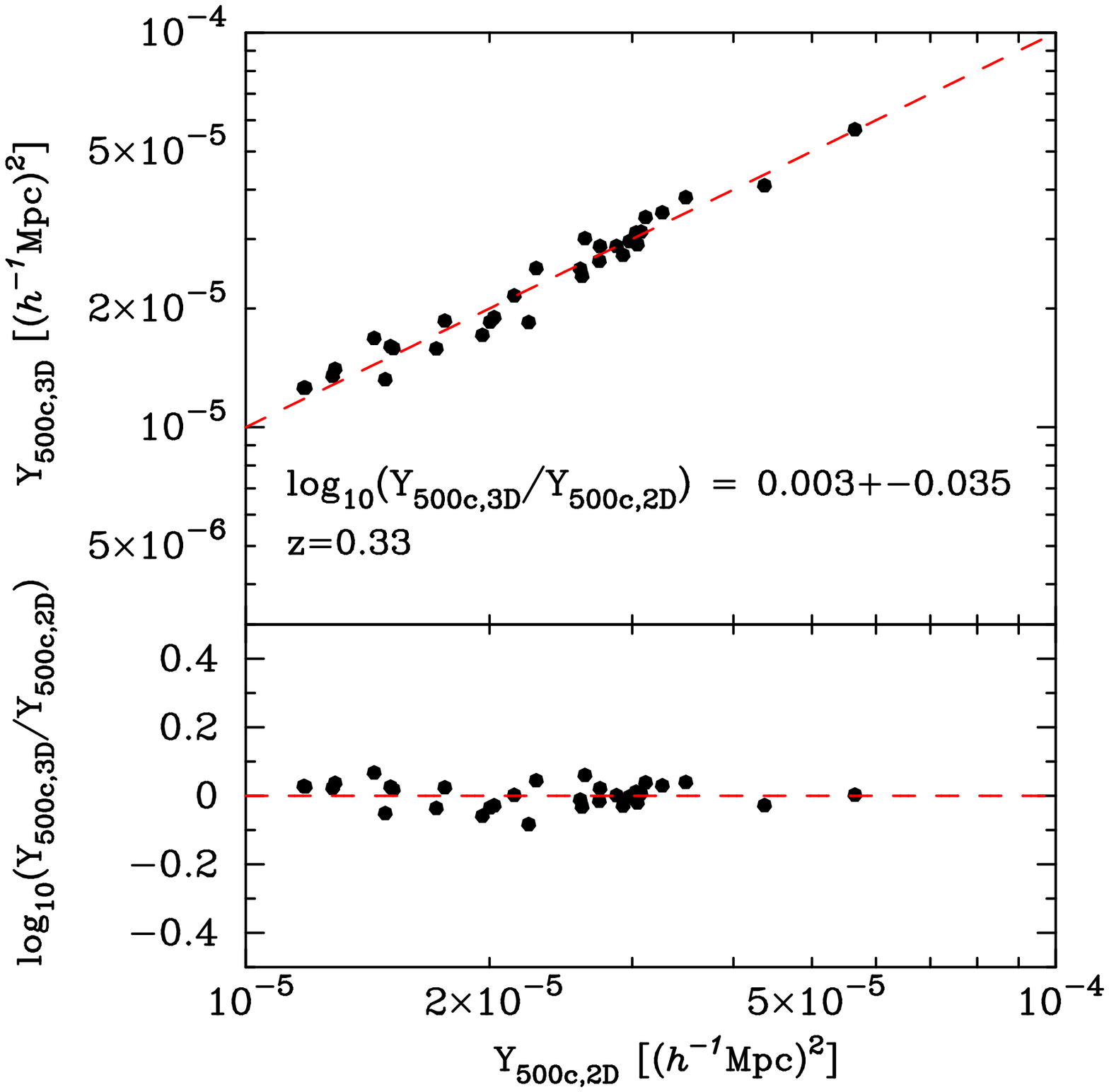}
\includegraphics[clip, width=1.0\columnwidth]{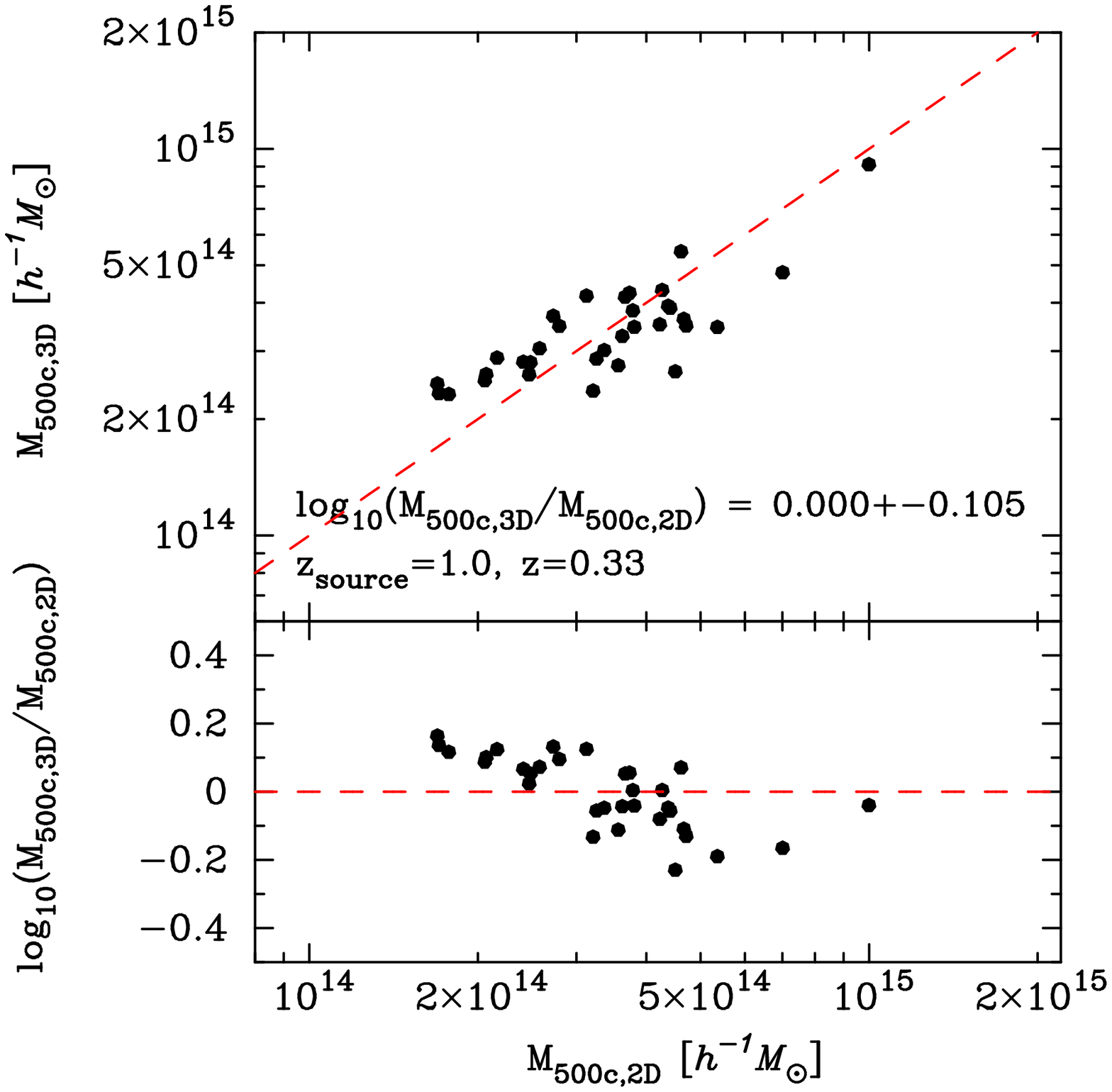}
\caption{
	The comparison of $Y_{\rm 2D}$ and $Y_{\rm 3D}$ 
	({\it left panel}) and $M_{\rm 2D}$ and $M_{\rm 3D}$ 
	({\it right panel}) for $33$ simulated clusters at 
	$z=0.33$ with the projection depth 
	$L_{\rm depth}=500\,h^{-1}\, {\rm Mpc}$ 
	for mock tSZ and WL maps.  
	The red dashed lines in the upper panels 
	represent one-to-one correspondence. 
	The bottom panels show the deviations of 
	the 2D measurements from the true 3D values. 
	In each panel, the numbers in the legend 
	indicate the average and the variance of 
	$\log (M_{\rm 3D}/M_{\rm 2D})$ or 
	$\log (Y_{\rm 3D}/Y_{\rm 2D)}$ over $33$ clusters. 
	\label{fig:compare_2D_3D}
	}
\end{figure*}

To understand the origin of the increased scatter, Figure~\ref{fig:compare_2D_3D} 
compares $Y_{\rm 2D}$ with $Y_{\rm 3D}$ and $M_{\rm 2D}$ with $M_{\rm 3D}$. 
The left panel shows the differences between $Y_{\rm 2D}$ and $Y_{\rm 3D}$, 
which shows that the relation between $Y_{\rm 2D}$ and $Y_{\rm 3D}$ is unbiased on average
with the scatter of $0.035$ in $\log Y_{\rm 2D}/Y_{\rm 3D}$. 
The right panel shows that the differences between $M_{\rm 2D}$ and $M_{\rm 3D}$. 
The scatter is relatively large ($\sim 0.105$), 
and the ratio of $M_{\rm 2D}/M_{\rm 3D}$ vs.\ $M_{\rm 2D}$ exhibits a ``tilt'', 
suggesting that $M_{\rm 2D}$ is a biased estimator of $M_{\rm 3D}$. 

We find that this ``tilt'' originates from the non-uniform distribution of the 
underlying true mass $M_{\rm 3D}$. 
If $M_{\rm 3D}$ does not follow a uniform distribution, 
which is the case for our simulated cluster sample, the mean value of 
$M_{\rm 2D}$ for a given $M_{\rm 3D}$ will be different from
the mean value of $M_{\rm 3D}$ for a given $M_{\rm 2D}$. 
Following the Appendix in \citet{2014MNRAS.438...62R}, 
the mean value of $\log M_{\rm 3D}$ for a given $\log M_{\rm 2D}$ is given by
\beqa
\langle \log M_{\rm 3D}|\log M_{\rm 2D} \rangle
= \log M_{\rm 2D} -\beta\sigma_{\log M, {\rm 2D}}^2,
\label{eq:model_Malm_bias}
\eeqa
where $\sigma_{\log M, {\rm 2D}}$ 
is the scatter in $\log M_{\rm 2D}$, and we assumed that 
the mean value of $\log M_{\rm 2D}$ is unbiased: 
$\langle \log M_{\rm 2D}|\log M_{\rm 3D} \rangle = \log M_{\rm 3D}$,
and the distribution of $m=\log M_{\rm 3D}$ can be expressed locally in $m$ as an exponential function $dn/dm \propto \exp(-\beta m)$. 
Thus, any non-zero $\beta$ and 
non-zero scatter in $\log M_{\rm 2D}$ gives rise to 
bias in $\langle \log M_{\rm 3D}|\log M_{\rm 2D} \rangle$.
Since our cluster sample is mass-limited, 
a sharp cut in the mass distribution can induce $\beta<0$ at 
$M\approx M_{\rm thre}=2.3\times10^{14}\, h^{-1}\, M_{\odot}$, 
while $\beta>0$ should hold 
at high-mass end where the mass function decreases exponentially.
Thus, the trend in the right bottom panel 
in Figure~\ref{fig:compare_2D_3D}
is consistent with the local model of the Malmquist bias 
\citep{2010MNRAS.408.1818W, 2010ApJ...715.1508S, 2014MNRAS.438...62R}, 
highlighting the importance of 
understanding the selection function of the observed 
cluster samples and correcting the Malmquist bias.

\subsection{Source of scatter in $Y_{\rm 2D}$ and $M_{\rm 2D}$}

\subsubsection{Projection effect}
\label{sec:proj}

Projection of line-of-sight structures is one of the primary sources of scatter in $Y_{\rm 2D}$ and $M_{\rm 2D}$
\citep[e.g.,][]{2007ApJ...671...27H, 2010A&A...514A..93M, 2012ApJ...758...74B}.
To quantify this effect, we compute $Y_{\rm 2D}$ from the tSZ maps using the four different projection depths 
$L_{\rm depth}=10, 20, 100$, and $500\, h^{-1}{\rm Mpc}$.
The pressure profile fitting is performed in the angular range of 
$0.1'$ to $\theta_{\rm 500c}$, where $\theta_{\rm 500c}$ is 
the angle corresponding to $R_{\rm 500c}$. In this section, we compute 
$Y_{\rm 2D}$ within the {\it true} cluster radius $R_{\rm 3D}$. 
Note, however, that the uncertainty in the halo radius 
$R_{\rm 500c}$ can introduce additional 
scatter, which will be examined separately in Section~\ref{sec:r_2d}. 

We quantify the scatter between $Y_{\rm 2D}$ and $Y_{\rm 3D}$ for our sample of $33$ clusters as
\beqa
\sigma_{\log  Y, {\rm 2D-3D}}^2 = \frac{1}{N_{\rm m}-1} \sum_{i=1}^{N_{\rm m}}\left( \log Y_{{\rm 2D}, i}-\log Y_{{\rm 3D},i}\right)^2,
\label{eq:y_2d_scatter}
\eeqa
where $Y_{{\rm 3D}, i}$ and $Y_{{\rm 2D}, i}$ are the 3D and 2D integrated Compton-$y$ values of the $i$-th cluster and $N_{\rm m}=33$.

\begin{table*}
\begin{tabular}{@{}ccc|ccc|}
\hline
\hline
$L_{\rm depth} [h^{-1}\, {\rm Mpc}]$ & $0.1'-5'$ & $0.1'-\theta_{\rm 500c}$ & $0.1'-5'$ & $0.1'-\theta_{\rm 500c}$ \\ \hline
$x$-axis projection & mass-limited sample & & without the outlier \\ \hline
$10$   
& $(3.91\pm0.04) \times 10^{-2}$ 
& $(2.74\pm0.02) \times 10^{-2}$ 
& $(2.97\pm0.02) \times 10^{-2}$ 
& $(2.53\pm0.02) \times 10^{-2}$ \\
$20$   
& $(3.39\pm0.03) \times 10^{-2}$ 
& $(2.64\pm0.02) \times 10^{-2}$ 
& $(3.10\pm0.02) \times 10^{-2}$ 
& $(2.65\pm0.02) \times 10^{-2}$ \\
$100$ 
& $(3.39\pm0.03) \times 10^{-2}$ 
& $(2.74\pm0.02) \times 10^{-2}$
& $(3.31\pm0.03) \times 10^{-2}$ 
& $(2.76\pm0.02) \times 10^{-2}$ \\
$500$ 
& $(3.50\pm0.03) \times 10^{-2}$ 
& $(2.80\pm0.02) \times 10^{-2}$
& $(3.40\pm0.03) \times 10^{-2}$ 
& $(2.83\pm0.02) \times 10^{-2}$ \\
\hline
$y$-axis projection & mass-limited sample & & without the outlier \\ \hline
$10$   
& $(3.25\pm0.03) \times 10^{-2}$ 
& $(2.29\pm0.01) \times 10^{-2}$
& $(2.75\pm0.02) \times 10^{-2}$ 
& $(2.10\pm0.01) \times 10^{-2}$ \\
$20$   
& $(3.40\pm0.03) \times 10^{-2}$ 
& $(2.38\pm0.02) \times 10^{-2}$ 
& $(2.90\pm0.02) \times 10^{-2}$ 
& $(2.18\pm0.01) \times 10^{-2}$ \\
$100$ 
& $(3.70\pm0.03) \times 10^{-2}$ 
& $(2.49\pm0.02) \times 10^{-2}$
& $(3.23\pm0.03) \times 10^{-2}$ 
& $(2.30\pm0.01) \times 10^{-2}$ \\
$500$ 
& $(3.94\pm0.04) \times 10^{-2}$ 
& $(2.63\pm0.02) \times 10^{-2}$ 
& $(3.45\pm0.03) \times 10^{-2}$ 
& $(2.43\pm0.02) \times 10^{-2}$ \\
\hline
$z$-axis projection & mass-limited sample & & without the outlier \\ \hline
$10$   
& $(3.88\pm0.04) \times 10^{-2}$ 
& $(2.65\pm0.02) \times 10^{-2}$ 
& $(2.87\pm0.02) \times 10^{-2}$ 
& $(2.34\pm0.01) \times 10^{-2}$ \\
$20$  
& $(4.18\pm0.04) \times 10^{-2}$ 
& $(2.94\pm0.02) \times 10^{-2}$ 
& $(3.22\pm0.03) \times 10^{-2}$ 
& $(2.65\pm0.02) \times 10^{-2}$ \\
$100$ 
& $(4.34\pm0.05) \times 10^{-2}$ 
& $(3.05\pm0.02) \times 10^{-2}$
& $(3.43\pm0.03) \times 10^{-2}$ 
& $(2.76\pm0.02) \times 10^{-2}$ \\
$500$ 
& $(4.40\pm0.05) \times 10^{-2}$ 
& $(3.10\pm0.03) \times 10^{-2}$ 
& $(3.52\pm0.03) \times 10^{-2}$ 
& $(2.82\pm0.02) \times 10^{-2}$ \\
\hline
\end{tabular}
\caption{
	The scatter between $\log Y_{{\rm 2D}}(R_{{\rm 3D}})$ 
	and $\log Y_{{\rm 3D}}(R_{{\rm 3D}})$ measured within the true $R_{\rm 500c}$. 
	The error is estimated by the Gaussian error over $33$ maps. 
	To convert the values into the conventional definition of scatter,
	multiply them by $\ln 10 \approx 2.3$. 
	The left portion shows the results for the mass-limited sample of $33$ clusters, while
	the right corresponds to the results for $32$ clusters without the $7\sigma$ outlier.
	\label{tab:scat_Y2D_R3D}
}
\end{table*}

Table~\ref{tab:scat_Y2D_R3D} shows how the projection effect introduces additional scatter in $Y_{2D}$ 
relative to the intrinsic scatter in $Y_{3D}$ as we increase the projection depth, $L_{\rm depth}$. 
For all three projections, we find a general trend that the scatter increases monotonically 
with $L_{\rm depth}$ from $20\,h^{-1}\, {\rm Mpc}$ to $500\,h^{-1}\, {\rm Mpc}$, except for 
one case between $L_{\rm depth}=10-20\,h^{-1}\, {\rm Mpc}$ in the $x$-axis projection.
In the case of $L_{\rm depth}=10\, h^{-1}\, {\rm Mpc}$, 
we find a cluster with $\log Y_{\rm 2D}/Y_{\rm 3D} \sim -0.2$, 
making this $7\sigma$ outlier in the population. 
We confirm that this is a merging cluster with 
$M_{\rm 500c}=4.1\times10^{14}\, h^{-1}M_{\odot}$ 
at $z=0.33$. The projected Compton-$y$ profile of this cluster 
has a flat core at $\theta < 1'$, which makes the gNFW model a poor fit.
We find that this merging cluster affects the estimation of scatter up to $30\%$ 
(see the right portion in Table~\ref{tab:scat_Y2D_R3D} for the result 
without the outlier).
When removing this cluster, the scatter increases monotonically 
with $L_{\rm depth}$ as expected in absence of such an outlier.
We also find that the scatter in the fitting range of $0.1'-\theta_{\rm 500c}$ is consistently 
smaller than the scatter based on the fitting range of $0.1'-5'$ for any given $L_{\rm depth}$,  
suggesting that the tSZ signal from $\theta\gtrsim \theta_{\rm 500c}$ is responsible for the additional scatter 
in $Y_{\rm 2D}(R_{\rm 3D})$.
We find that the scatter in $M_{\rm 2D}$ increases monotonically with the projection depth, 
$L_{\rm depth}$ \citep{2003MNRAS.339.1155H, 2004PhRvD..70b3008D, 2011MNRAS.412.2095H}, 
because of the increased contribution from the uncorrelated matter distribution along 
the line of sight \citep{2015MNRAS.449.4264G}.
Moreover, the scatter in $\ln M_{\rm 2D}$ with $L_{\rm depth}=500\, h^{-1}{\rm Mpc}$ 
is $0.24$, which is similar to that in the scatter of 
$0.22$ reported in the previous study based on a large cosmological 
N-body simulation with a box size of $1\, h^{-1}\, {\rm Gpc}$ \citep{2011ApJ...740...25B}. 

\subsubsection{Uncertainties in estimated halo radius from WL}
\label{sec:r_2d}

\begin{figure}
\centering 
\includegraphics[clip, width=1.0\columnwidth]{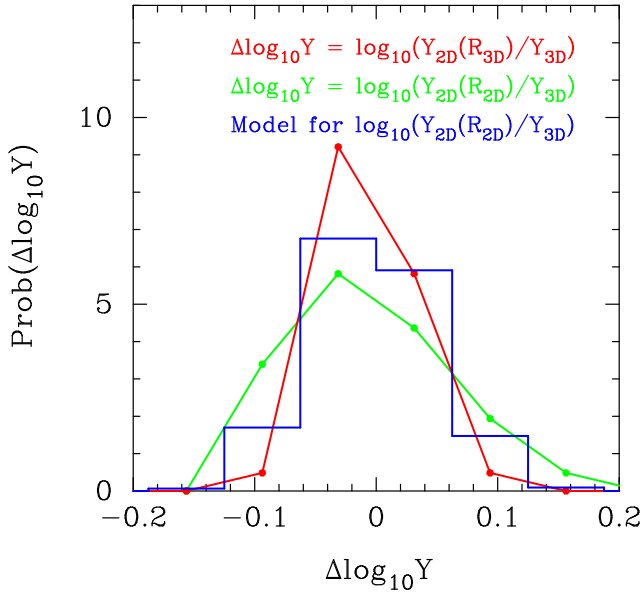}
\caption{
	The probability distribution of $\Delta \log Y=\log(Y_{\rm 2D}(R_{\rm 3D})/Y_{\rm 3D})$ 
	(red line) and $\Delta \log Y=\log(Y_{\rm 2D}(R_{\rm 2D})/Y_{\rm 3D})$ (green line). 
	The blue histogram corresponds to our modeling with 
	the log-normal distribution of 
	$\log \left(M_{\rm 2D}/M_{\rm 3D}\right)$.
	\label{fig:dist_dlogY}
}
\end{figure}

Another major source of the scatter in $Y_{\rm 2D}$ is the biased estimation of the halo 
radius $R_{\rm 2D}$ resulting from the bias in the WL mass, which enters into our 
calculation of the integrated $Y$ in Eq.~(\ref{eq:def_Y}).  

To quantify this effect, we compare 
$Y_{\rm 2D}(R_{\rm 2D})/Y_{\rm 3D}$ with $Y_{\rm 2D}(R_{\rm 3D})/Y_{\rm 3D}$, 
where $Y_{\rm 2D}(R_{\rm 3D})$ and $Y_{\rm 2D}(R_{\rm 2D})$ are computed 
within the true radius $R_{\rm 3D}$ and the WL estimated radius $R_{\rm 2D}$, respectively.
In both cases, we compute the scatter in $Y_{\rm 2D}/Y_{\rm 3D}$ using Eq.~(\ref{eq:y_2d_scatter}).

Figure~\ref{fig:dist_dlogY} shows the distribution function of 
the deviation of projected $Y_{\rm 2D}$ from the
true $Y_{\rm 3D}$ for $33$ clusters obtained from the WL and tSZ maps with 
the projection depth of $L_{\rm depth}=500\,h^{-1}\, {\rm Mpc}$ along the $x$-axis. 
The red line represents the distribution where the projected $Y_{\rm 2D}$
is measured within $R_{\rm 3D}$ 
(i.e., $\Delta \log Y(R_{\rm 3D})=\log \left(Y_{\rm 2D}(R_{\rm 3D})/Y_{\rm 3D}\right)$),
while the green line shows the distribution where the projected $Y_{\rm 2D}$
is measured within $R_{\rm 2D}$ estimated from the WL mass 
(i.e., $\Delta \log Y(R_{\rm 2D})=\log \left(Y_{\rm 2D}(R_{\rm 2D})/Y_{\rm 3D}\right)$).
The distribution of $\Delta \log Y(R_{\rm 2D})$ is broader than that of $\Delta \log Y(R_{\rm 3D})$, 
indicating that WL mass measurements of $M_{\rm 2D}$ introduce additional scatter 
in $Y_{\rm 2D}$ by $11.0\%$, which is larger than $4.5\%$ increase in scatter due to 
projection effects discussed in the Section~\ref{sec:proj}. Note that similar results are 
obtained for the other two projection axes, where the additional scatters in $Y_{\rm 2D}$ 
are found to be $9.0\%$ and $10.4\%$ for $y$-axis and $z$-axis, respectively.
This shows that the uncertainty in $M_{\rm 2D}$ leads to significant scatter in 
the WL calibration of the $Y-M$ relations, and this effect must be taken into account in 
the cosmological parameter estimation based on WL mass calibration of SZ-selected cluster samples.

In order to account for this effect, 
we develop a model to predict the distribution of $\log \left(Y_{\rm 2D}(R_{\rm 2D})/Y_{\rm 3D}\right)$
for a given $\log \left(Y_{\rm 2D}(R_{\rm 3D})/Y_{\rm 3D}\right)$.
Assuming that the underlying pressure profile is given by the gNFW pressure profile with the best-fit 
parameters $M_{500,p}$ and $c_{500,p}$, the uncertainty in WL mass $M_{\rm 2D}$ is translated into 
the uncertainty in $R_{\rm ref}$ through Eq.~(\ref{eq:def_Y}). 
Note that the integral in Eq.~(\ref{eq:def_Y}) scales with the following quantity:
\beqa
I_{\rm P}(x_{\rm out}) = \int_{0}^{x_{\rm out}} p(x) 4\pi x^2{\rm d}x,
\eeqa
where $x_{\rm out} = R_{\rm ref}/R_{500,p}$, $p(x)$ is given by Eq.~(\ref{eq:gNFW}) and 
$\alpha^{\prime}_{\rm P}$ is assumed to play a minor role in the evaluation of this integral. 
Since $\delta \log  I_{\rm P} \simeq \delta \log  x_{\rm out}$ at $x_{\rm out}=1$,
the uncertainty in $\log  M_{\rm 2D}$ introduces the scatter in $\log  Y_{\rm 2D}$ 
by $\delta \log  R_{\rm 2D} \sim \delta \log  M_{\rm 2D}/3$.
We can then model the probability distribution of 
$\log Y_1 = \log \left(Y_{\rm 2D}(R_{\rm 2D})\right)$ based on 
the probability distribution of $\log Y_2 =\log \left(Y_{\rm 3D}(R_{\rm 3D})\right)$ as 
\beqa
\wp(\log Y_1) = \int {\rm d}\log Y_2\, \wp(\log Y_2)\, \wp(\log Y_1|\log Y_2),
\label{eq:model_dlogY}
\eeqa
where $\wp(\log Y_1|\log Y_2)$ is 
the distribution of $\log Y_1$ for a given $\log Y_2$.
We assume $\wp(\log Y_1|\log Y_2)$ to be the log-normal distribution with the scatter of 
$(1/3)\, \sigma_{\log M {\rm 2D}-{\rm 3D}}$, where $\sigma_{\log M {\rm 2D}-{\rm 3D}}$ 
is the scatter of $\log \left(M_{\rm 2D}/M_{\rm 3D}\right)$.
The blue histogram in Figure~\ref{fig:dist_dlogY} is the result of our model, which provides
a good description of our simulation results.

\section{Covariance between tSZ and WL signals}
\label{sec:covariance}

\begin{figure}
\centering \includegraphics[clip, width=1.0\columnwidth]{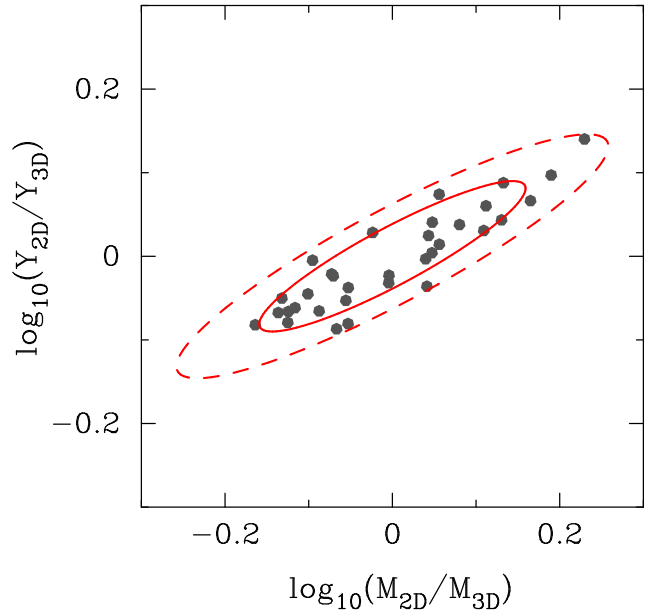}
\caption{
	The covariance between $M_{\rm 2D}$ and $Y_{\rm 2D}$
	derived from WL mass and tSZ maps of $33$ simulated clusters.
	The gray point shows the scatter plot of 
	$\log \left(M_{\rm 2D}/M_{\rm 3D}\right)$
	and $\log \left(Y_{\rm 2D}/Y_{\rm 3D}\right)$. 
	The red solid and dashed lines indicate 
	the $1\sigma$ and $2\sigma$ contours of 
	the two-dimensional log-normal distribution 
	with the measured covariance, respectively.
	\label{fig:cov_M2D_Y2D}
	}
\end{figure}

The scatter in $Y_{\rm 2D}$ is likely correlated with the scatter in $M_{\rm 2D}$,
as they are both affected by the projection effects and the uncertainties in the estimation of $M_{\rm 2D}$. 
Therefore, the covariance between $Y_{\rm 2D}$ and $M_{\rm 2D}$ must be taken 
into account in order to derive the unbiased estimate of the underlying $Y_{\rm 3D}-M_{\rm 3D}$ 
relations from tSZ and WL measurements. 

\subsection{Covariance in the $Y-M$ relation}
\label{sec:cov}

In order to characterize the nature of scatter in the observed $Y_{\rm 2D}-M_{\rm 2D}$ 
scaling relation, we quantify the correlation between the scatters in 
$Y_{\rm 2D}$ and $M_{\rm 2D}$ with the covariance matrix ${\bd C}$ of the two-dimensional variable
${\bd X} = ({\log (M_{\rm 2D}/M_{\rm 3D})}, \log (Y_{\rm 2D}/Y_{\rm 3D}))$
as follows:
\beqa
C_{ij}  
&=& \frac{1}{N_{\rm m}-1}
\sum_{k=1}^{N_{\rm m}}(X_{ki}-\bar{X}_{i})(X_{kj}-\bar{X}_{j}), \\ \label{eq:def_cov}
{\bar X}_{i} 
&=& \frac{1}{N_{\rm m}}
\sum_{k=1}^{N_{\rm m}}X_{ki}, \label{eq:def_av}
\eeqa
where $X_{ki}$ represents the $i$-th component of ${\bd X}$ for the $k$-th map.

The resulting covariance matrix for the $33$ simulated clusters viewed along the $x$ projection axis is
\beqa
{\bd C}=
\left(
\begin{array}{cc}
1.12\times10^{-2} & 5.67\times10^{-3} \\
5.67\times10^{-3} & 3.55\times10^{-3} \\
\end{array}
\right). \label{eq:cov_M2D_Y2D}
\eeqa
Figure~\ref{fig:cov_M2D_Y2D} shows the covariance
between $Y_{\rm 2D}$ and $M_{\rm 2D}$
for the $33$ simulated clusters viewed along the $x$ projection axis, where
the grey points represent the resulting ${\bd X}$ from a $\chi^2$ fitting, 
and the red lines are the $1\sigma$ and $2\sigma$ contours of the log-normal distribution with 
the covariance matrix ${\bd C}$ in Eq.~(\ref{eq:cov_M2D_Y2D}).  
The points trace the log-normal contours quite well.
We also find that the scatter in $\log (M_{\rm 2D}/M_{\rm 3D})$
is tightly correlated with that of $\log (Y_{\rm 2D}/Y_{\rm 3D})$. 
The correlation coefficients for our simulated clusters are 
$0.902$, $0.769$ and $0.828$ for the $x, y, z$ projection axes, respectively.
Removing the outlier discussed in Section~\ref{sec:proj} changes 
the correlation coefficient by $\simlt0.02$.
The significant covariance between the scatter in 
$Y_{\rm 2D}$ and $M_{\rm 2D}$ we found
is consistent with previous theoretical studies on covariance between
cluster observables 
\citep{2010MNRAS.408.1818W, 2010ApJ...715.1508S, 2012MNRAS.426.2046A, 2012MNRAS.426.1829N}
and observational work \citep[e.g.,][]{2009ApJ...699..768R}.

Another important correlation in the tSZ and WL measurement 
is the covariance between $M_{\rm 2D}$ and $Y_{\rm 2D}$
at a given $M_{\rm 3D}$.
This covariance ${\bd C}^{\prime}$ 
is defined by the two-dimensional variable
of ${\bd X}^{\prime}=(\log(M_{\rm 2D}/M_{\rm 3D}),
\log(Y_{\rm 2D}/Y_{\rm 3D, scal}))$, 
where $Y_{\rm 3D, scal}$ is given by Eq.~(\ref{eq:true_Y-M})
at a given $M_{\rm 3D}$.
For the 33 simulated clusters viewed along the x projection axis,
we found that
\beqa
{\bd C}^{\prime}=
\left(
\begin{array}{cc}
1.12\times10^{-2} & 6.29\times10^{-3} \\
6.29\times10^{-3} & 4.45\times10^{-3} \\
\end{array}
\right). \label{eq:cov_M2D_Y2D_alt}
\eeqa
Compared to Eq.~(\ref{eq:cov_M2D_Y2D}), the scatter 
in $\log(Y_{\rm 2D}/Y_{\rm 3D, scal})$ is larger 
than that in $\log(Y_{\rm 2D}/Y_{\rm 3D})$ 
because of the scatter in $Y_{\rm 3D, scal}$, 
while the correlation coefficient changes only by $\sim0.01$.
Similar results are found for the other two projection axes.

\subsection{Recovering the unbiased 3D $Y-M$ Relation}
\label{sec:reconstruct_3D}

With the covariance between $Y_{\rm 2D}$ and $M_{\rm 2D}$ in hand, we can develop a statistical 
model to recover the underlying $Y_{\rm 3D}-M_{\rm 3D}$ relation from a set of measurements of 
($M_{\rm 2D}$,$Y_{\rm 2D}$) using the bayesian framework as follows. 

Let the distribution of true halo mass $M_{\rm 3D}$ to be $\wp(M_{\rm 3D})$ for 
WL mass ranging between $M_{\rm 2D}$ and $M_{\rm 2D}+{\rm d}M_{\rm 2D}$
and $Y_{\rm 2D}$ ranging between $Y_{\rm 2D}$ and $Y_{\rm 2D}+{\rm d}Y_{\rm 2D}$. 
The differential number density of the cluster haloes is then given by
\beqa
\frac{{\rm d}N\left(M_{\rm 2D}, Y_{\rm 2D}\right)}{{\rm d}M_{\rm 2D}{\rm d}Y_{\rm 2D}} =\int {\rm d}Y_{\rm 3D}{\rm d}M_{\rm 3D}\times \nonumber \\ 
\wp(M_{\rm 3D}) \wp(Y_{\rm 3D}|M_{\rm 3D})
\wp(M_{\rm 2D}, Y_{\rm 2D}|M_{\rm 3D}, Y_{\rm 3D}),
\label{eq:model_M2D_Y2D}
\eeqa
where $\wp(Y_{\rm 3D}|M_{\rm 3D})$ represents the probability distribution of the underlying $Y_{\rm 3D}-M_{\rm 3D}$ relation and $\wp(M_{\rm 2D}, Y_{\rm 2D}|M_{\rm 3D}, Y_{\rm 3D})$
is the probability distribution function of a set of $(M_{\rm 2D}, Y_{\rm 2D})$ for a
given set of $(M_{\rm 3D}, Y_{\rm 3D})$.
Assuming that they follow the log-normal distributions, we have 
\beqa
\wp(Y_{\rm 3D}|M_{\rm 3D}) = A \exp \left\{ -\frac{1}{2}\left[
\frac{\log Y_{\rm 3D}-\log  Y_{\rm model}}{\sigma}\right]^2\right\}, 
\label{eq:prob_Y3D_M3D} 
\eeqa
where $A=1/\sqrt{2\pi\sigma^2}$, $\sigma=\sigma_{\log  Y, {\rm 3D}}$ and 
$\log  Y_{\rm model}=\alpha_{0}+\alpha_{1}\log \left(M_{\rm 3D}/(10^{14} h^{-1}\, M_{\odot})\right)$, and 
\beqa
\wp(M_{\rm 2D}, Y_{\rm 2D}|M_{\rm 3D}, Y_{\rm 3D}) =B\exp \left\{
-\frac{1}{2}{\bd X}^{T} {\bd C}^{-1} {\bd X}
\right\}, \label{eq:prob_Y2D_M3D_Y3D_M3D}
\eeqa
where ${\bd X} = (\log (M_{\rm 2D}/M_{\rm 3D}), \log (Y_{\rm 2D}/Y_{\rm 3D}))$, $B=1/\sqrt{(2\pi)^2\det {\bd C}}$, and ${\bd C}$ represents the covariance matrix of ${\bd X}$. 

\begin{figure}
\centering \includegraphics[clip, width=0.95\columnwidth]{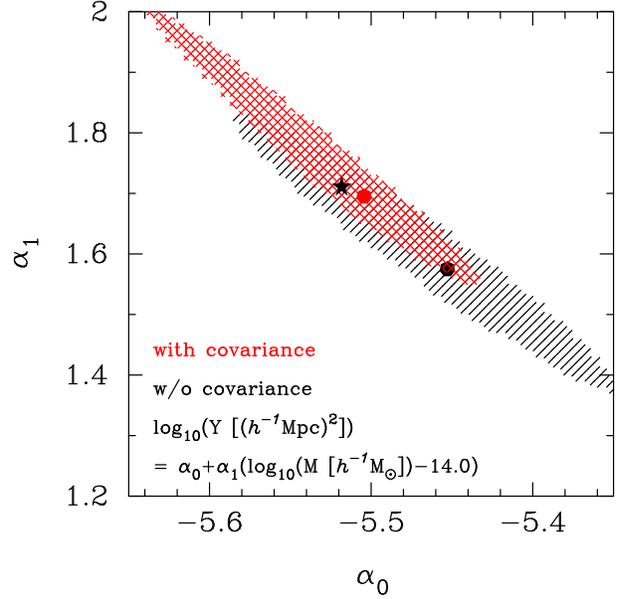}
\caption{
	The posterior distribution of the parameters of the $Y_{\rm 2D}-M_{\rm 2D}$ relation
	for $33$ simulated clusters.
	The red filled circle shows the best-fit parameters derived from 
	the likelihood analysis with the covariance between $M_{\rm 2D}$ 
	and $Y_{\rm 2D}$.
	The black filled circle is the best-fit parameters
	when $Y_{\rm 2D}$ and $M_{\rm 2D}$ are 
	assumed to be independent,
	while the black star symbol represents the best-fit parameters 
	of the $Y_{\rm 3D}-M_{\rm 3D}$ relation.
	The hatched region shows the 95\% confidence level of
	the posterior distribution. 
	\label{fig:likelihood}
	}
\end{figure}

Figure~\ref{fig:scat_M2D_Y2D} shows that our model is able to recover the $Y_{\rm 2D}-M_{\rm 2D}$ scaling relation, 
with the true scaling relation $\wp(Y_{\rm 3D}|M_{\rm 3D})$ 
and the covariance ${\bd C}$ measured from our simulation. 
The red points show the expected distribution of the model and the best-fit parameters 
$\alpha_{0}, \alpha_{1}, \sigma_{\log Y, {\rm 3D}}$ and ${\bd C}$. The red error bars represent the $68\%$ 
confidence level of $\log Y_{\rm 2D}$ for a given $\log M_{\rm 2D}$. 
The red points recover our 2D measurements indicated by grey points, 
demonstrating that our model provides a good description of the $Y_{\rm 2D}-M_{\rm 2D}$ relation from tSZ-WL mock analyses.
We stress that the covariance is an essential ingredient
in explaining the scatter in the $Y_{\rm 2D}-M_{\rm 2D}$ relation
in Figure~\ref{fig:scat_M2D_Y2D}.
The scatter of  $\sim14\%$ in
$\log (Y_{\rm 2D}/Y_{\rm 3D})$ alone 
is not enough to explain the total scatter of $\sim23\%$.
One also have to 
include the covariance between 
$\log (Y_{\rm 2D}/Y_{\rm 3D})$ and
$\log (M_{\rm 2D}/Y_{\rm 3D})$. 

Next, we recover the $Y_{\rm 3D}-M_{\rm 3D}$ relation from our model by estimating 
the parameters $\alpha_{0}$ and $\alpha_{1}$ in Eq.~(\ref{eq:prob_Y3D_M3D}). 
To do this, we first construct the likelihood function of number density of clusters 
in the $Y_{\rm 2D}-M_{\rm 2D}$ assuming the Poisson distribution:
\beqa
{\cal L} = \prod_{i}^{N_{\log Y}}\prod_{j}^{N_{\log M}}\frac{\lambda^{N_{ij}}\exp(-\lambda)}{N_{ij}!},
\label{eq:likelihood}
\eeqa
where $N_{ij}$ is the number count of clusters found in $(i,j)$-th grid in the 
$Y_{\rm 2D}-M_{\rm 2D}$ plane, 
$N_{\log Y}$ and $N_{\log M}$ represent the number of bins in 
$\log  Y_{\rm 2D}$ and $\log  M_{\rm 2D}$, respectively.
The best-fit parameters $\alpha_{0}$ and $\alpha_{1}$ are then found by maximizing 
the likelihood ${\cal L}$.
We test our method with measured values of $Y_{\rm 2D}$ and $M_{\rm 2D}$ 
over $33\times3=99$ realizations of projected cluster maps (by combing simulated 
clusters viewed along three orthogonal projections) 
with $L_{\rm depth}=500\, h^{-1}\, {\rm Mpc}$.
The likelihood function is calculated over 100 logarithmically space bins in 
$10^{14}<M_{\rm 2D} \, [h^{-1}\, M_{\odot}]<10^{15}$
and $10^{-5.5}< Y_{\rm 2D} \, [(h^{-1}\,{\rm Mpc})^2]<10^{-4}$. 
For simplicity, we set $\sigma_{\log Y, {\rm 3D}}=0.030$ and adopt the distribution of $M_{\rm 3D}$ measured from our simulations
(see the black hatched histogram in Figure~\ref{fig:dist_mass_3d}).

The result of our likelihood analysis is summarized in Figure~\ref{fig:likelihood}.
The black star symbol represents the parameters of 
the underlying 3D $Y-M$ relation. 
The red point is for the best-fit parameters obtained from our likelihood analysis. 
The red hatched region shows the 95\% confidence level of the posterior distribution 
of $\alpha_{0}$ and $\alpha_{1}$.
The true parameters is well within the red hatched region, demonstrating 
that our maximum likelihood analysis 
can recover the true 3D scaling relation reasonably well.
We emphasize that it is critical to include the covariance 
$\bd C$ between $\log (Y_{\rm 2D}/Y_{\rm 3D})$ 
and $\log (M_{\rm 2D}/M_{\rm 3D})$. 
Ignoring it leads to biases in the estimated parameters 
of the 3D scaling relation, as illustrated by the black point and hatched region in Figure~\ref{fig:likelihood}.
Note that the bias in the estimated slope ($\alpha_1$) 
of the $Y-M$ relation is on the order $\sim0.10$, 
which is comparable to the statistical uncertainty in the current observations
\citep[e.g.,][]{2015arXiv150201597P,2016arXiv160306522D}.
Thus, the covariance among cluster observables must be taken into account in order to take advantage
of the statistical power of current and future tSZ and WL cluster surveys. 

After recovering the unbiased 3D $Y-M$ relation, one can reduce the uncertainty in 
the estimate of $Y_{\rm 2D}(R_{\rm 2D})$ by an iterative approach as follows 
(see also \cite{2015MNRAS.448.2085L}).
Using the $Y_{\rm 3D}-M_{\rm 3D}$ relation, 
one can compute a new 
estimate of $M_{\rm 3D}=f(Y_{\rm 3D})$ to re-define the boundary of a cluster 
$R_{\rm 3D}$ through $M_{\rm 500c} = 500 \rho_{\rm crit}(z) (4\pi/3)R_{\rm 500c}^3$.
One can then iterate to obtain a new estimate of $Y_{\rm 2D}$ within the new radius $R_{\rm 3D}$. 
This iterative approach is expected to be efficient 
because the scatter in WL mass is larger than the scatter in $Y$ at a given $M_{\rm 3D}$.
We tested this iterative approach by using the mock measurements of $Y_{\rm 2D}$ and 
the $Y_{\rm 3D}-M_{\rm 3D}$ relation in Eq.~(\ref{eq:true_Y-M}).
In the case of $L_{\rm proj}=500\, h^{-1}{\rm Mpc}$,
we found that the scatter in $\log (Y_{\rm 2D}/Y_{\rm 3D})$ 
changes from 5.9\% to 4.5\% for $x$-axis after ten iterations, which 
was sufficient for convergence of results.
Note that similar results are also obtained for the other two axes,
where the scatter decreases from 5.6\% to 5.1\% and
from 6.3\% to 5.5\% for $y$-axis and $z$-axis, respectively.
While this iterative approach is useful to obtain a more accurate estimate of $Y_{\rm 2D}$, 
it still does not completely remove the uncertainty in $R_{\rm 2D}$ in measurement of $Y_{\rm 2D}$; 
i.e., we cannot reduce the scatter of $\log(Y_{\rm 2D}(R_{\rm 2D})/Y_{\rm 3D})$ to that of 
$\log(Y_{\rm 2D}(R_{\rm 3D})/Y_{\rm 3D})$ through this iterative approach.

\subsection{Implications for Cosmological Inferences}
\label{sec:cosmology}

Finally, we assess the impact of the biased $Y-M$ relation 
on cluster-based cosmological constraints. Here, we consider 
the cumulative number count of galaxy clusters as a
function of the angular integrated Compton-$y$ parameter 
$Y_{\rm ang} = 1/D^2_{A}(z) Y$,
where $D_{A}(z)$ is the angular diameter distance for redshift of $z$.
The number count per solid angle in the redshift range of 
$z_{\rm min}$ to $z_{\rm max}$ is given by
\beqa
&&N(Y_{\rm ang,thre}; z_{\rm min}, z_{\rm max}) = \nonumber \\ 
&& \int_{z_{\rm min}}^{z_{\rm max}} {\rm d}z \, \frac{{\rm d}V}{{\rm d}z}
\int_{Y_{\rm ang,thre}(M,z)}^{\infty} {\rm d} M \, 
\frac{{\rm d}n}{{\rm d}M}\, \wp(Y_{\rm ang}|M, z),
\label{eq:number_count}
\eeqa
where ${\rm d}n/{\rm d}M$ is the halo mass function and $\wp(Y_{\rm ang}|M, z)$ 
expresses the scaling relation between $Y_{\rm ang}$ and mass $M$ at redshift $z$.
We use the halo mass function by \citet{2008ApJ...688..709T}, and $\wp(Y_{\rm ang}|M, z)$ is set to 
be the log-normal function with the scatter of 0.18 \citep{2012MNRAS.426.2046A}. 
As a fiducial model, we consider the self-similar $Y-M$ relation 
as shown in Eq~\ref{eq:true_Y-M}
with cosmological parameters set to the {\em WMAP} nine-year results \citep{Hinshaw2013}.
We consider two additional scenarios where the $Y-M$ relation is biased when the covariance
between the scatters in $Y$ and $M$ are ignored, as shown by the black point in Figure~\ref{fig:likelihood}.  
In one scenario, we set our cosmological parameters to the fiducial WMAP9 values, 
while in the other we increase $\Omega_{\rm m0}$ higher by 2.5\%, which corresponds to 
the $1\sigma$ error in the WMAP9 value.
Note that we take into account changes in both halo mass function and 
angular diameter distance when varying cosmological parameters.

\begin{figure}
\centering \includegraphics[clip, width=1.0\columnwidth]{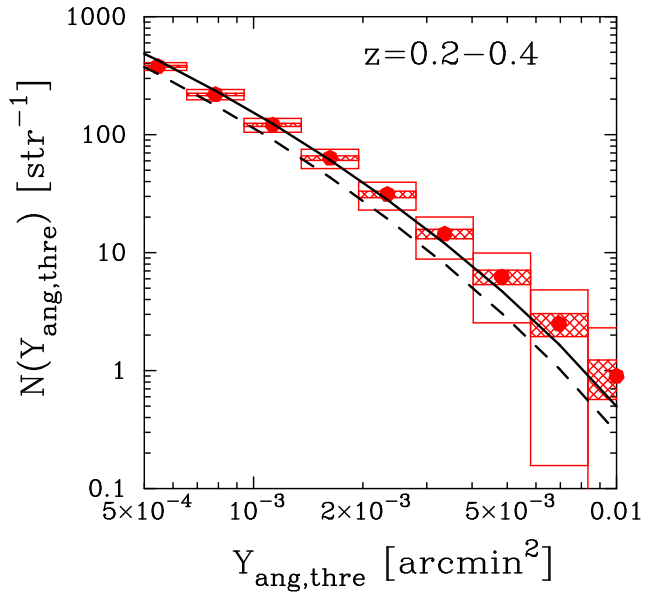}
\caption{
	Number count of galaxy clusters at $z=0.2-0.4$ as a function 
	of the angular integrated Compton-$y$ parameter ($Y_{\rm ang}$).
	The red points represent the expected value with the self-similar scaling relation.
	The open and hatched boxes indicate the Poisson errors for the sky coverage of 
	$1,500$ and $27,000$ squared degrees, respectively.
	The black dashed line corresponds to the model for the biased $Y-M$ relation 
	in the WMAP9 cosmology, while the black solid line is the prediction based on 
	the biased $Y-M$ relation and cosmology with 
	higher matter density $\Omega_{\rm m0}$ by $2.5$\%. 
	\label{fig:impact_numbercount}
	}
\end{figure}

For illustration, we consider the redshift range of $z=0.2-0.4$, which is the relevant redshift
range for recent WL measurements of tSZ-selected clusters 
\citep[e.g.,][]{2012ApJ...758...68H, 2015arXiv150908930B}.
Figure~\ref{fig:impact_numbercount} shows the expected cluster number counts 
for the three different models. 
The red points represent our fiducial case, the black dashed line corresponds to 
the biased $Y-M$ relation with the fiducial cosmology,
and the black solid line corresponds to the case with the biased $Y-M$ relation 
and with higher $\Omega_{\rm m0}$. 
The red open and hatched boxes show the Poisson error for a hypothetical survey
with the sky coverage of 1,500 and 27,000 squared degrees, 
which correspond to the coverage of ongoing imaging surveys (such as the Hyper Suprime-Cam) 
and the full-sky coverage with masking of the galactic plane, respectively. 
For a fixed cosmology, the biased $Y-M$ relation leads to reduction in 
the number count in the survey area of 27,000 squared degrees, which is comparable to 
the sample size of the {\em Planck} tSZ cluster catalog \citep{2014A&A...571A..29P, 2015arXiv150201598P}.
Increasing $\Omega_{\rm m0}$ leads to higher cluster counts, suggesting that the biased $Y-M$ relation 
can introduce biases in cosmological parameters, such as $\Omega_{\rm m0}$ and $\sigma_8$. In this case,
10\% bias in the $Y-M$ relation leads to an increase of 2.5\% in $\Omega_{\rm m0}$, or an increase of 6.6\% in $\sigma_{8}$ for a fixed initial curvature perturbation amplitude.

\begin{table*}
\begin{tabular}{@{}ccc|ccc|}
\hline
\hline
$L_{\rm depth} [h^{-1}\, {\rm Mpc}]$ & $0.1'-5'$ & $0.1'-\theta_{\rm 500c}$ & $0.1'-5'$ & $0.1'-\theta_{\rm 500c}$ \\ \hline
$x$-axis projection & mass-limited sample & & without the outliers \\ \hline
$10$   
& $(5.87\pm0.07) \times 10^{-2}$ 
& $(5.27\pm0.06) \times 10^{-2}$ 
& $(4.80\pm0.05) \times 10^{-2}$ 
& $(3.81\pm0.03) \times 10^{-2}$ \\
$20$   
& $(6.88\pm0.10) \times 10^{-2}$ 
& $(5.81\pm0.07) \times 10^{-2}$ 
& $(5.21\pm0.06) \times 10^{-2}$ 
& $(4.02\pm0.04) \times 10^{-2}$ \\
$100$ 
& $(6.90\pm0.10) \times 10^{-2}$ 
& $(5.83\pm0.07) \times 10^{-2}$
& $(5.23\pm0.06) \times 10^{-2}$ 
& $(4.05\pm0.04) \times 10^{-2}$ \\
$500$ 
& $(6.94\pm0.10) \times 10^{-2}$ 
& $(5.83\pm0.07) \times 10^{-2}$
& $(5.34\pm0.06) \times 10^{-2}$ 
& $(4.11\pm0.04) \times 10^{-2}$ \\
\hline
$y$-axis projection & mass-limited sample & & without the outliers \\ \hline
$10$   
& $(3.96\pm0.03) \times 10^{-2}$ 
& $(3.23\pm0.03) \times 10^{-2}$
& $(4.02\pm0.03) \times 10^{-2}$ 
& $(3.28\pm0.02) \times 10^{-2}$ \\
$20$   
& $(4.10\pm0.04) \times 10^{-2}$ 
& $(3.36\pm0.03) \times 10^{-2}$ 
& $(4.16\pm0.04) \times 10^{-2}$ 
& $(3.41\pm0.03) \times 10^{-2}$ \\
$100$ 
& $(4.25\pm0.04) \times 10^{-2}$ 
& $(3.43\pm0.03) \times 10^{-2}$
& $(4.31\pm0.04) \times 10^{-2}$ 
& $(3.48\pm0.03) \times 10^{-2}$ \\
$500$ 
& $(4.37\pm0.04) \times 10^{-2}$ 
& $(3.48\pm0.03) \times 10^{-2}$ 
& $(4.42\pm0.04) \times 10^{-2}$ 
& $(3.53\pm0.03) \times 10^{-2}$ \\
\hline
$z$-axis projection & mass-limited sample & & without the outliers \\ \hline
$10$   
& $(4.45\pm0.04) \times 10^{-2}$ 
& $(3.54\pm0.03) \times 10^{-2}$ 
& $(4.03\pm0.03) \times 10^{-2}$ 
& $(2.92\pm0.02) \times 10^{-2}$ \\
$20$  
& $(4.60\pm0.05) \times 10^{-2}$ 
& $(3.84\pm0.03) \times 10^{-2}$ 
& $(4.19\pm0.04) \times 10^{-2}$ 
& $(3.27\pm0.02) \times 10^{-2}$ \\
$100$ 
& $(4.65\pm0.05) \times 10^{-2}$ 
& $(3.87\pm0.03) \times 10^{-2}$
& $(4.23\pm0.04) \times 10^{-2}$ 
& $(3.30\pm0.02) \times 10^{-2}$ \\
$500$ 
& $(4.86\pm0.05) \times 10^{-2}$ 
& $(4.02\pm0.03) \times 10^{-2}$ 
& $(4.42\pm0.04) \times 10^{-2}$ 
& $(3.43\pm0.03) \times 10^{-2}$ \\
\hline
\end{tabular}
\caption{
	Same as Table~\ref{tab:scat_Y2D_R3D}, but for the CSF run. 
	The left portion shows the results for the mass-limited 
	sample of $46$ clusters, while
	the right portion corresponds to the results of $44$ clusters 
	without the two outliers described in the text.
	\label{tab:scat_Y2D_R3D_CSF}
}
\end{table*}

\begin{figure}
\centering \includegraphics[clip, width=1.0\columnwidth]{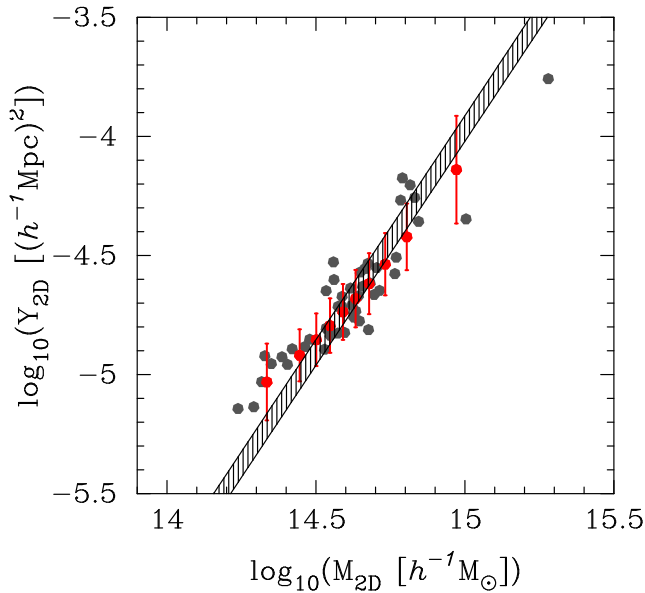}
\caption{
	The scatter plot of the $Y-M$ relation inferred 
	from two-dimensional tSZ and WL mock observations.
	The gray points represent the result of $Y_{\rm 2D}$ 
	and $M_{\rm 2D}$
	obtained from a $\chi^2$ fitting over 46 realizations 
	of tSZ and WL maps.
	The black hatched region corresponds to 
	the underlying $Y-M$ relation 
	for our simulated clusters with the scatter of 
	$\sigma_{\log Y} = 0.050$,
	or $\sigma_{\ln Y} = \sigma_{\log Y} \times \ln 10 = 11.5\%$. 
	The red point with error bar shows our modeling of the 
	$Y_{\rm 2D}-M_{\rm 2D}$ relation.
	\label{fig:scat_M2D_Y2D_CSF}
	}
\end{figure}

\section{Baryonic Effects}
\label{sec:baryon}

So far, the simulations we have treat the ICM as a non-radiative gas and ignored additional baryonic physics, 
such as radiative cooling, star formation, and feedback from active galactic nuclei. These baryonic physics can in principle 
induce additional scatter in the observed $Y_{\rm 2D}-M_{\rm 2D}$ relation by changing the level of gas pressure in 
the correlated structure along the line of sight. While these effects are expected to be small 
\citep{2006ApJ...650..538N,2012ApJ...758...74B,2012MNRAS.422.1999K}, further scrutiny is still useful in order 
to assess to what extent the impact of the uncertain baryonic physics on the $Y_{\rm 2D}-M_{\rm 2D}$ relation.

In order to examine the effects of baryonic physics on the scatter of $Y-M$ relation, we analyzed 
re-simulation of {\em Omega500} with radiative cooling, star formation, and supernova feedback (CSF). 
This CSF run includes metallicity-dependent radiative cooling,
star formation, thermal supernova feedback, metal enrichment and advection, 
which are based on the same subgrid physics modules in \citet{2007ApJ...668....1N},
which we refer the reader for more details. 
In the following, we work with a mass-limited sample of 46 clusters 
with $M_{\rm 500c}\geq2.8\times10^{14}\, h^{-1} M_{\odot}$ at $z=0.33$.
Note that our CSF simulation suffers from the well-known ``overcooling'' problem, where the simulation
over-predicts the amount of central stellar mass by a factor of $\sim 2$. As such, the results of our NR 
and CSF run can be used to {\em bracket} systematic uncertainties associated with baryonic effects.

Following the analyses in Section~\ref{sec:scatter}, we first measure the $Y-M$ relation 
and its intrinsic scatter in the CSF run. We find that the best-fit scaling relation between 
$\log Y_{\rm 3D}$ and $\log M_{\rm 3D}$ is
\beqa
\log \left(\frac{Y_{\rm 3D}}{(h^{-1}\,{\rm Mpc})^2}\right)= 
1.88\log \left(\frac{M_{\rm 3D}}{10^{14}\,h^{-1}\,M_{\odot}}\right)
-5.84,
\label{eq:true_Y-M_CSF}
\eeqa
where the best-fit slope of $1.88\pm 0.030$ ($1\sigma$ error; see the best-fit
relation shown as the hatched region in Figure~\ref{fig:scat_M2D_Y2D_CSF}) 
is different from the self-similar prediction of $5/3$ because of the increasingly 
larger reduction in the gas mass fraction at the low-mass clusters \citep[e.g.,][]{2006ApJ...650..538N}. 
The intrinsic scatter in the CSF run is $\sigma_{\log Y, {\rm 3D}}=0.050$, 
suggesting that gas cooling and star formation can increase the intrinsic
scatter of the $Y_{\rm 3D}-M_{\rm 3D}$ relation by up to $70\%$. 
We find that the increased scatter originates from the enhanced fluctuations 
in gas pressure in the CSF run relative to the NR run 
\citep[see also][]{2013MNRAS.431..954K}. Note that the scatter changes 
by only $\simlt4\%$ when excising the core region ($R\leq0.15R_{500c}$), 
concluding that the cluster core makes a minor contribution to the scatter. 

Table~\ref{tab:scat_Y2D_R3D_CSF} reports the scatter between 
$Y_{\rm 2D}$ and $Y_{\rm 3D}$ in the CSF run. 
Analogous to the NR case, we find that the scatter of the CSF run increases 
with the projection depth $L_{\rm proj}$ from 10 to 500 $h^{-1}{\rm Mpc}$ for three different projections. 
For $L_{\rm proj}=500\, h^{-1}{\rm Mpc}$ and fitting range of 0.1'-5',
we find that the baryonic effects change the scatter 
between $Y_{\rm 2D}$ and $Y_{\rm 3D}$ by about $10\%$,
except for the $x$-axis projection.
In the $x$-axis projection, we find two clusters with 
$\log Y_{\rm 2D}/Y_{\rm 3D}\sim0.3$ and $-0.5$, making them 
$6\sigma$ and $7\sigma$ outliers in the population, respectively.
The $6\sigma$ outlier has two high-pressure cores within $R_{\rm 500c}$.
One of the cores is located around $\theta_{\rm 500c}$ in the projected 
Compton-$y$ map, causing a poor gNFW model fit. The $7\sigma$ outlier 
has a flat core at $\theta<1'$, making the gNFW a poor fit. When removing 
these outliers, there is a clearer trend of increasing scatter with $L_{\rm proj}$. 

Finally, we measure the covariance between tSZ and WL signals to be
\beqa
{\bd C}=
\left(
\begin{array}{cc}
1.33\times10^{-2} & 1.06\times10^{-3} \\
1.06\times10^{-2} & 1.22\times10^{-2} \\
\end{array}
\right), \label{eq:cov_M2D_Y2D_CSF}
\eeqa
for the x-axis projection in the CSF run. We also confirmed that the two-dimensional variable
${\bd X} = ({\log (M_{\rm 2D}/M_{\rm 3D})}, \log (Y_{\rm 2D}/Y_{\rm 3D}))$ 
follows the bivariate Gaussian distribution with the covariance matrix for the CSF run.
The correlation coefficients are found to be $0.838$, $0.706$ and $0.690$ for 
the $x, y, z$ projections, respectively, which differ from the NR values at the level of $\simlt20\%$. 

In summary, baryonic effects can alter the statistical property of tSZ and WL 
signals at some level. 
However, we show that our model can accommodate the 
baryonic effects, by taking into account changes in the $Y_{\rm 3D}-M_{\rm 3D}$ relation, 
its intrinsic scatter, and the covariance matrix of the two-dimensional variable, 
${\bd X}=(\log(M_{\rm 2D}/M_{\rm 3D}),\log(Y_{\rm 2D}/Y_{\rm 3D}))$. 
In Figure~\ref{fig:scat_M2D_Y2D_CSF}, the gray points show 
the measured $Y_{\rm 2D}$ and $M_{\rm 2D}$ of the CSF clusters, and 
the red points with error bar represent our modeling as shown in Section~\ref{sec:reconstruct_3D}.
Our model shows that the scatter in the WL calibrated $Y-M$ relation 
is $28\%$ in the CSF run, compared to $23\%$ in the NR run. Since the NR and CSF runs should bracket the range of baryonic effects, 
we expect that the 
realistic model should lie within the range explored in this work. 

\section{CONCLUSIONS}
\label{sec:conclusions}

The tSZ effect is widely recognized as a robust mass proxy of galaxy clusters with 
small intrinsic scatter. However, recent observational calibration of the tSZ-WL mass 
relation shows that the observed scatter is considerably larger than the intrinsic scatter predicted by numerical simulations.
This raises a question as to whether we can exploit the full statistical power of upcoming SZ and WL cluster surveys. 
In this work, we investigated the origin of observed scatter in the $Y-M$ relations, 
using mock tSZ and WL maps of galaxy clusters extracted from high-resolution cosmological hydrodynamical simulations. 
Our main findings are summarized as follows: 

\begin{enumerate}

\vspace{2mm}
\item We showed that the scatter in the WL calibrated $Y-M$ relation is $23\%$. This is significantly larger than 
the intrinsic scatter of $\lesssim 10\%$ predicted by simulations, and it is consistent with the observed scatter of about $20\%$.  

\vspace{2mm}
\item The uncertainty in the integrated Compton-y, $Y$, inferred from the projected Compton-$y$ profile originates 
from the combination of (a) the projection effect in the tSZ maps and (b) the uncertainty in the cluster radius determined 
from the WL mass measurements, with each effect contributing to the total scatter by 5\% and 10\%, respectively. 

\vspace{2mm}
\item
The scatter in the tSZ-WL mass relation can be explained by the combination of uncertainties associated with 
$Y$ and WL mass measurements. Namely, the amplitude of the scatter is determined by the {\em covariance} between tSZ and WL signals. 
In the presence of the uncertainty in the WL mass, the distribution of clusters in the $Y_{\rm 2D}-M_{\rm 2D}$ 
plane is smeared in both $Y_{\rm 2D}$ and $M_{\rm 2D}$, where its scatter is different from the scatter in $\log Y_{\rm 2D}$ alone.

\vspace{2mm}
\item
We show that the covariance between tSZ and WL signals is important for recovering the true $Y-M$ relation. 
Ignoring the covariance would lead to 10\% bias in the $Y-M$ relation, which leads to the biases
in $\Omega_{\rm m0}$ by 2.5\%, and $\sigma_{8}$ by 6.6\%. 
Thus, this covariance must be taken into account for cosmological constraints with ongoing and future cluster surveys. 

\vspace{2mm}
\item
We show that the covariance of the $Y-M$ relation depends on the input baryonic physics at a level of $\simlt20\%$, 
by using two sets of simulations that bracket a broad range of astrophysical uncertainties. We further demonstrate 
that our statistical model to describe the $Y_{\rm 2D}-M_{\rm 2D}$ relation can provide a reasonable description of 
the simulation results, provided that the proper modeling of the true $Y-M$ relation and 
the covariance in the $Y_{\rm 2D}-M_{\rm 2D}$ plane are performed. 

\vspace{2mm}
\item We present a statistical model to recover the unbiased $Y-M$ relation from a set of tSZ and WL measurements 
which enables us to obtain the unbiased tSZ-mass scaling relation from a simultaneous measurement of tSZ and WL, 
and opens up the possibility of extracting cosmological information from upcoming multi-wavelength surveys that will 
provide a large statistical sample of galaxy clusters out to the high-redshift ($z\simlt 1$) universe. 

\end{enumerate}

Future work should focus on developing and analyzing a larger sample of simulated clusters and tSZ and WL mocks 
in order to characterize the mass and redshift dependence of the $Y_{\rm 2D}-M_{\rm 2D}$ relation, its covariance 
matrix, and the impact of the outlier populations due to mergers. Addressing these issues is the critical step for 
understanding the remaining astrophysical uncertainties and hence accurate and robust interpretations of current 
and upcoming SZ and lensing surveys, including cluster counts, SZ power spectrum and higher-order moments, 
and cross-correlations between tSZ and WL maps.


\section*{acknowledgments} 

We thank Nick Battaglia, Nikhel Gupta, Eduardo Rozo, Alex Saro, Hironao Miyatake, and the anonymous 
referee for comments on the manuscript. We also acknowledge the Max-Planck-Institut f{\"u}r Astrophysik 
for their hospitality during the workshop ``ICM Physics and Modeling'' (2015). 
MS is supported by Research Fellowships of the Japan Society for the Promotion of Science (JSPS) 
for Young Scientists. DN and EL acknowledge support from NSF grant AST-1412768, NASA ATP grant NNX11AE07G, 
NASA Chandra Theory grant GO213004B, the Research Corporation, and by the facilities and staff of the Yale Center for Research Computing.


\bibliographystyle{mn2e}
\bibliography{bibtex_library}

\begin{thebibliography}{63}
\expandafter\ifx\csname natexlab\endcsname\relax\def\natexlab#1{#1}\fi

\bibitem[{{Angulo} {et~al}\mbox{.}(2012){Angulo}, {Springel}, {White},
  {Jenkins}, {Baugh}, \& {Frenk}}]{2012MNRAS.426.2046A}
{Angulo} R.~E., {Springel} V., {White} S.~D.~M., {Jenkins} A., {Baugh} C.~M.,
  {Frenk} C.~S., 2012, MNRAS, 426, 2046

\bibitem[{{Arnaud} {et~al}\mbox{.}(2010){Arnaud}, {Pratt}, {Piffaretti},
  {B{\"o}hringer}, {Croston}, \& {Pointecouteau}}]{2010A&A...517A..92A}
{Arnaud} M., {Pratt} G.~W., {Piffaretti} R., {B{\"o}hringer} H., {Croston}
  J.~H., {Pointecouteau} E., 2010, A\&A, 517, A92

\bibitem[{Bartelmann \& Schneider(2001)}]{Bartelmann2001}
Bartelmann M., Schneider P., 2001, Physics Reports, 340, 291

\bibitem[{{Battaglia} {et~al}\mbox{.}(2012){Battaglia}, {Bond}, {Pfrommer}, \&
  {Sievers}}]{2012ApJ...758...74B}
{Battaglia} N., {Bond} J.~R., {Pfrommer} C., {Sievers} J.~L., 2012, ApJ, 758,
  74

\bibitem[{{Battaglia} {et~al}\mbox{.}(2015){Battaglia}, {Leauthaud},
  {Miyatake}, {Hasselfield}, {Gralla}, {Allison}, {Bond}, {Calabrese},
  {Crichton}, {Devlin}, {Dunkley}, {D{\"u}nner}, {Erben}, {Ferrara}, {Halpern},
  {Hilton}, {Hill}, {Hincks}, {Hlo{\v z}ek}, {Huffenberger}, {Hughes}, {Kneib},
  {Kosowsky}, {Makler}, {Marriage}, {Menanteau}, {Miller}, {Moodley}, {Moraes},
  {Niemack}, {Page}, {Shan}, {Sehgal}, {Sherwin}, {Sievers}, {Sif{\'o}n},
  {Spergel}, {Staggs}, {Taylor}, {Thornton}, {van Waerbeke}, \&
  {Wollack}}]{2015arXiv150908930B}
{Battaglia} N. {et~al.}, 2015, ArXiv e-prints

\bibitem[{{Becker} \& {Kravtsov}(2011)}]{2011ApJ...740...25B}
{Becker} M.~R., {Kravtsov} A.~V., 2011, ApJ, 740, 25

\bibitem[{{Bleem} {et~al}\mbox{.}(2015){Bleem}, {Stalder}, {de Haan}, {Aird},
  {Allen}, {Applegate}, {Ashby}, {Bautz}, {Bayliss}, {Benson}, {Bocquet},
  {Brodwin}, {Carlstrom}, {Chang}, {Chiu}, {Cho}, {Clocchiatti}, {Crawford},
  {Crites}, {Desai}, {Dietrich}, {Dobbs}, {Foley}, {Forman}, {George},
  {Gladders}, {Gonzalez}, {Halverson}, {Hennig}, {Hoekstra}, {Holder},
  {Holzapfel}, {Hrubes}, {Jones}, {Keisler}, {Knox}, {Lee}, {Leitch}, {Liu},
  {Lueker}, {Luong-Van}, {Mantz}, {Marrone}, {McDonald}, {McMahon}, {Meyer},
  {Mocanu}, {Mohr}, {Murray}, {Padin}, {Pryke}, {Reichardt}, {Rest}, {Ruel},
  {Ruhl}, {Saliwanchik}, {Saro}, {Sayre}, {Schaffer}, {Schrabback},
  {Shirokoff}, {Song}, {Spieler}, {Stanford}, {Staniszewski}, {Stark}, {Story},
  {Stubbs}, {Vanderlinde}, {Vieira}, {Vikhlinin}, {Williamson}, {Zahn}, \&
  {Zenteno}}]{2015ApJS..216...27B}
{Bleem} L.~E. {et~al.}, 2015, ApJS, 216, 27

\bibitem[{{Bocquet} {et~al}\mbox{.}(2015){Bocquet}, {Saro}, {Mohr}, {Aird},
  {Ashby}, {Bautz}, {Bayliss}, {Bazin}, {Benson}, {Bleem}, {Brodwin},
  {Carlstrom}, {Chang}, {Chiu}, {Cho}, {Clocchiatti}, {Crawford}, {Crites},
  {Desai}, {de Haan}, {Dietrich}, {Dobbs}, {Foley}, {Forman}, {Gangkofner},
  {George}, {Gladders}, {Gonzalez}, {Halverson}, {Hennig}, {Hlavacek-Larrondo},
  {Holder}, {Holzapfel}, {Hrubes}, {Jones}, {Keisler}, {Knox}, {Lee}, {Leitch},
  {Liu}, {Lueker}, {Luong-Van}, {Marrone}, {McDonald}, {McMahon}, {Meyer},
  {Mocanu}, {Murray}, {Padin}, {Pryke}, {Reichardt}, {Rest}, {Ruel}, {Ruhl},
  {Saliwanchik}, {Sayre}, {Schaffer}, {Shirokoff}, {Spieler}, {Stalder},
  {Stanford}, {Staniszewski}, {Stark}, {Story}, {Stubbs}, {Vanderlinde},
  {Vieira}, {Vikhlinin}, {Williamson}, {Zahn}, \&
  {Zenteno}}]{2015ApJ...799..214B}
{Bocquet} S. {et~al.}, 2015, ApJ, 799, 214

\bibitem[{{de Haan} {et~al}\mbox{.}(2016){de Haan}, {Benson}, {Bleem}, {Allen},
  {Applegate}, {Ashby}, {Bautz}, {Bayliss}, {Bocquet}, {Brodwin}, {Carlstrom},
  {Chang}, {Chiu}, {Cho}, {Clocchiatti}, {Crawford}, {Crites}, {Desai},
  {Dietrich}, {Dobbs}, {Doucouliagos}, {Foley}, {Forman}, {Garmire}, {George},
  {Gladders}, {Gonzalez}, {Gupta}, {Halverson}, {Hlavacek-Larrondo},
  {Hoekstra}, {Holder}, {Holzapfel}, {Hou}, {Hrubes}, {Huang}, {Jones},
  {Keisler}, {Knox}, {Lee}, {Leitch}, {von der Linden}, {Luong-Van}, {Mantz},
  {Marrone}, {McDonald}, {McMahon}, {Meyer}, {Mocanu}, {Mohr}, {Murray},
  {Padin}, {Pryke}, {Rapetti}, {Reichardt}, {Rest}, {Ruel}, {Ruhl},
  {Saliwanchik}, {Saro}, {Sayre}, {Schaffer}, {Schrabback}, {Shirokoff},
  {Song}, {Spieler}, {Stalder}, {Stanford}, {Staniszewski}, {Stark}, {Story},
  {Stubbs}, {Vanderlinde}, {Vieira}, {Vikhlinin}, {Williamson}, \&
  {Zenteno}}]{2016arXiv160306522D}
{de Haan} T. {et~al.}, 2016, ArXiv e-prints

\bibitem[{{Dodelson}(2004)}]{2004PhRvD..70b3008D}
{Dodelson} S., 2004, Physical Review D, 70, 023008

\bibitem[{{Gruen} {et~al}\mbox{.}(2015){Gruen}, {Seitz}, {Becker}, {Friedrich},
  \& {Mana}}]{2015MNRAS.449.4264G}
{Gruen} D., {Seitz} S., {Becker} M.~R., {Friedrich} O., {Mana} A., 2015, MNRAS,
  449, 4264

\bibitem[{{Gruen} {et~al}\mbox{.}(2014){Gruen}, {Seitz}, {Brimioulle},
  {Kosyra}, {Koppenhoefer}, {Lee}, {Bender}, {Riffeser}, {Eichner},
  {Weidinger}, \& {Bierschenk}}]{2014MNRAS.442.1507G}
{Gruen} D. {et~al.}, 2014, MNRAS, 442, 1507

\bibitem[{{Hallman} {et~al}\mbox{.}(2007){Hallman}, {O'Shea}, {Burns},
  {Norman}, {Harkness}, \& {Wagner}}]{2007ApJ...671...27H}
{Hallman} E.~J., {O'Shea} B.~W., {Burns} J.~O., {Norman} M.~L., {Harkness} R.,
  {Wagner} R., 2007, ApJ, 671, 27

\bibitem[{{Hasselfield} {et~al}\mbox{.}(2013){Hasselfield}, {Hilton},
  {Marriage}, {Addison}, {Barrientos}, {Battaglia}, {Battistelli}, {Bond},
  {Crichton}, {Das}, {Devlin}, {Dicker}, {Dunkley}, {D{\"u}nner}, {Fowler},
  {Gralla}, {Hajian}, {Halpern}, {Hincks}, {Hlozek}, {Hughes}, {Infante},
  {Irwin}, {Kosowsky}, {Marsden}, {Menanteau}, {Moodley}, {Niemack}, {Nolta},
  {Page}, {Partridge}, {Reese}, {Schmitt}, {Sehgal}, {Sherwin}, {Sievers},
  {Sif{\'o}n}, {Spergel}, {Staggs}, {Swetz}, {Switzer}, {Thornton}, {Trac}, \&
  {Wollack}}]{2013JCAP...07..008H}
{Hasselfield} M. {et~al.}, 2013, JCAP, 7, 8

\bibitem[{{High} {et~al}\mbox{.}(2012){High}, {Hoekstra}, {Leethochawalit}, {de
  Haan}, {Abramson}, {Aird}, {Armstrong}, {Ashby}, {Bautz}, {Bayliss}, {Bazin},
  {Benson}, {Bleem}, {Brodwin}, {Carlstrom}, {Chang}, {Cho}, {Clocchiatti},
  {Conroy}, {Crawford}, {Crites}, {Desai}, {Dobbs}, {Dudley}, {Foley},
  {Forman}, {George}, {Gladders}, {Gonzalez}, {Halverson}, {Harrington},
  {Holder}, {Holzapfel}, {Hoover}, {Hrubes}, {Jones}, {Joy}, {Keisler}, {Knox},
  {Lee}, {Leitch}, {Liu}, {Lueker}, {Luong-Van}, {Mantz}, {Marrone},
  {McDonald}, {McMahon}, {Mehl}, {Meyer}, {Mocanu}, {Mohr}, {Montroy},
  {Murray}, {Natoli}, {Nurgaliev}, {Padin}, {Plagge}, {Pryke}, {Reichardt},
  {Rest}, {Ruel}, {Ruhl}, {Saliwanchik}, {Saro}, {Sayre}, {Schaffer}, {Shaw},
  {Schrabback}, {Shirokoff}, {Song}, {Spieler}, {Stalder}, {Staniszewski},
  {Stark}, {Story}, {Stubbs}, {{\v S}uhada}, {Tokarz}, {van Engelen},
  {Vanderlinde}, {Vieira}, {Vikhlinin}, {Williamson}, {Zahn}, \&
  {Zenteno}}]{2012ApJ...758...68H}
{High} F.~W. {et~al.}, 2012, ApJ, 758, 68

\bibitem[{Hinshaw {et~al}\mbox{.}(2013)Hinshaw, Larson, Komatsu, Spergel,
  Bennett, Dunkley, Nolta, Halpern, Hill, Odegard, Page, Smith, Weiland, Gold,
  Jarosik, Kogut, Limon, Meyer, Tucker, Wollack, \& Wright}]{Hinshaw2013}
Hinshaw G. {et~al.}, 2013, ApJS, 208, 19

\bibitem[{{Hoekstra}(2003)}]{2003MNRAS.339.1155H}
{Hoekstra} H., 2003, MNRAS, 339, 1155

\bibitem[{{Hoekstra} {et~al}\mbox{.}(2011){Hoekstra}, {Hartlap}, {Hilbert}, \&
  {van Uitert}}]{2011MNRAS.412.2095H}
{Hoekstra} H., {Hartlap} J., {Hilbert} S., {van Uitert} E., 2011, MNRAS, 412,
  2095

\bibitem[{{Hoekstra} {et~al}\mbox{.}(2012){Hoekstra}, {Mahdavi}, {Babul}, \&
  {Bildfell}}]{2012MNRAS.427.1298H}
{Hoekstra} H., {Mahdavi} A., {Babul} A., {Bildfell} C., 2012, MNRAS, 427, 1298

\bibitem[{{Jee} {et~al}\mbox{.}(2014){Jee}, {Hughes}, {Menanteau}, {Sif{\'o}n},
  {Mandelbaum}, {Barrientos}, {Infante}, \& {Ng}}]{2014ApJ...785...20J}
{Jee} M.~J., {Hughes} J.~P., {Menanteau} F., {Sif{\'o}n} C., {Mandelbaum} R.,
  {Barrientos} L.~F., {Infante} L., {Ng} K.~Y., 2014, ApJ, 785, 20

\bibitem[{{Kay} {et~al}\mbox{.}(2012){Kay}, {Peel}, {Short}, {Thomas}, {Young},
  {Battye}, {Liddle}, \& {Pearce}}]{2012MNRAS.422.1999K}
{Kay} S.~T., {Peel} M.~W., {Short} C.~J., {Thomas} P.~A., {Young} O.~E.,
  {Battye} R.~A., {Liddle} A.~R., {Pearce} F.~R., 2012, MNRAS, 422, 1999

\bibitem[{{Khedekar} {et~al}\mbox{.}(2013){Khedekar}, {Churazov}, {Kravtsov},
  {Zhuravleva}, {Lau}, {Nagai}, \& {Sunyaev}}]{2013MNRAS.431..954K}
{Khedekar} S., {Churazov} E., {Kravtsov} A., {Zhuravleva} I., {Lau} E.~T.,
  {Nagai} D., {Sunyaev} R., 2013, MNRAS, 431, 954

\bibitem[{{Klypin} {et~al}\mbox{.}(2001){Klypin}, {Kravtsov}, {Bullock}, \&
  {Primack}}]{2001ApJ...554..903K}
{Klypin} A., {Kravtsov} A.~V., {Bullock} J.~S., {Primack} J.~R., 2001, ApJ,
  554, 903

\bibitem[{{Komatsu} {et~al}\mbox{.}(2009){Komatsu}, {Dunkley}, {Nolta},
  {Bennett}, {Gold}, {Hinshaw}, {Jarosik}, {Larson}, {Limon}, {Page},
  {Spergel}, {Halpern}, {Hill}, {Kogut}, {Meyer}, {Tucker}, {Weiland},
  {Wollack}, \& {Wright}}]{2009ApJS..180..330K}
{Komatsu} E. {et~al.}, 2009, ApJS, 180, 330

\bibitem[{{Krause} {et~al}\mbox{.}(2012){Krause}, {Pierpaoli}, {Dolag}, \&
  {Borgani}}]{2012MNRAS.419.1766K}
{Krause} E., {Pierpaoli} E., {Dolag} K., {Borgani} S., 2012, MNRAS, 419, 1766

\bibitem[{{Kravtsov}(1999)}]{1999PhDT........25K}
{Kravtsov} A.~V., 1999, PhD thesis, New Mexico State University

\bibitem[{{Kravtsov}, {Klypin} \& {Hoffman}(2002){Kravtsov}, {Klypin}, \&
  {Hoffman}}]{2002ApJ...571..563K}
{Kravtsov} A.~V., {Klypin} A., {Hoffman} Y., 2002, ApJ, 571, 563

\bibitem[{{Liu} {et~al}\mbox{.}(2015){Liu}, {Mohr}, {Saro}, {Aird}, {Ashby},
  {Bautz}, {Bayliss}, {Benson}, {Bleem}, {Bocquet}, {Brodwin}, {Carlstrom},
  {Chang}, {Chiu}, {Cho}, {Clocchiatti}, {Crawford}, {Crites}, {de Haan},
  {Desai}, {Dietrich}, {Dobbs}, {Foley}, {Gangkofner}, {George}, {Gladders},
  {Gonzalez}, {Halverson}, {Hennig}, {Hlavacek-Larrondo}, {Holder},
  {Holzapfel}, {Hrubes}, {Jones}, {Keisler}, {Lee}, {Leitch}, {Lueker},
  {Luong-Van}, {McDonald}, {McMahon}, {Meyer}, {Mocanu}, {Murray}, {Padin},
  {Pryke}, {Reichardt}, {Rest}, {Ruel}, {Ruhl}, {Saliwanchik}, {Sayre},
  {Schaffer}, {Shirokoff}, {Spieler}, {Stalder}, {Staniszewski}, {Stark},
  {Story}, {{\v S}uhada}, {Vanderlinde}, {Vieira}, {Vikhlinin}, {Williamson},
  {Zahn}, \& {Zenteno}}]{2015MNRAS.448.2085L}
{Liu} J. {et~al.}, 2015, MNRAS, 448, 2085

\bibitem[{{Marrone} {et~al}\mbox{.}(2012){Marrone}, {Smith}, {Okabe},
  {Bonamente}, {Carlstrom}, {Culverhouse}, {Gralla}, {Greer}, {Hasler},
  {Hawkins}, {Hennessy}, {Joy}, {Lamb}, {Leitch}, {Martino}, {Mazzotta},
  {Miller}, {Mroczkowski}, {Muchovej}, {Plagge}, {Pryke}, {Sanderson},
  {Takada}, {Woody}, \& {Zhang}}]{2012ApJ...754..119M}
{Marrone} D.~P. {et~al.}, 2012, ApJ, 754, 119

\bibitem[{{Marrone} {et~al}\mbox{.}(2009){Marrone}, {Smith}, {Richard}, {Joy},
  {Bonamente}, {Hasler}, {Hamilton-Morris}, {Kneib}, {Culverhouse},
  {Carlstrom}, {Greer}, {Hawkins}, {Hennessy}, {Lamb}, {Leitch}, {Loh},
  {Miller}, {Mroczkowski}, {Muchovej}, {Pryke}, {Sharp}, \&
  {Woody}}]{2009ApJ...701L.114M}
{Marrone} D.~P. {et~al.}, 2009, ApJL, 701, L114

\bibitem[{{McInnes} {et~al}\mbox{.}(2009){McInnes}, {Menanteau}, {Heavens},
  {Hughes}, {Jimenez}, {Massey}, {Simon}, \& {Taylor}}]{2009MNRAS.399L..84M}
{McInnes} R.~N., {Menanteau} F., {Heavens} A.~F., {Hughes} J.~P., {Jimenez} R.,
  {Massey} R., {Simon} P., {Taylor} A., 2009, MNRAS, 399, L84

\bibitem[{{Meneghetti} {et~al}\mbox{.}(2010){Meneghetti}, {Rasia}, {Merten},
  {Bellagamba}, {Ettori}, {Mazzotta}, {Dolag}, \&
  {Marri}}]{2010A&A...514A..93M}
{Meneghetti} M., {Rasia} E., {Merten} J., {Bellagamba} F., {Ettori} S.,
  {Mazzotta} P., {Dolag} K., {Marri} S., 2010, A\&A, 514, A93

\bibitem[{{Miyatake} {et~al}\mbox{.}(2013){Miyatake}, {Nishizawa}, {Takada},
  {Mandelbaum}, {Mineo}, {Aihara}, {Spergel}, {Bickerton}, {Bond}, {Gralla},
  {Hajian}, {Hilton}, {Hincks}, {Hughes}, {Infante}, {Lin}, {Lupton},
  {Marriage}, {Marsden}, {Menanteau}, {Miyazaki}, {Moodley}, {Niemack},
  {Oguri}, {Price}, {Reese}, {Sif{\'o}n}, {Wollack}, \&
  {Yasuda}}]{2013MNRAS.429.3627M}
{Miyatake} H. {et~al.}, 2013, MNRAS, 429, 3627

\bibitem[{{Motl} {et~al}\mbox{.}(2005){Motl}, {Hallman}, {Burns}, \&
  {Norman}}]{2005ApJ...623L..63M}
{Motl} P.~M., {Hallman} E.~J., {Burns} J.~O., {Norman} M.~L., 2005, ApJL, 623,
  L63

\bibitem[{Munshi {et~al}\mbox{.}(2008)Munshi, Valageas, Vanwaerbeke, \&
  Heavens}]{Munshi2008}
Munshi D., Valageas P., Vanwaerbeke L., Heavens a., 2008, Physics Reports, 462,
  67

\bibitem[{{Nagai}(2006)}]{2006ApJ...650..538N}
{Nagai} D., 2006, ApJ, 650, 538

\bibitem[{{Nagai}, {Kravtsov} \& {Vikhlinin}(2007){Nagai}, {Kravtsov}, \&
  {Vikhlinin}}]{2007ApJ...668....1N}
{Nagai} D., {Kravtsov} A.~V., {Vikhlinin} A., 2007, ApJ, 668, 1

\bibitem[{{Nagai}, {Vikhlinin} \& {Kravtsov}(2007){Nagai}, {Vikhlinin}, \&
  {Kravtsov}}]{2007ApJ...655...98N}
{Nagai} D., {Vikhlinin} A., {Kravtsov} A.~V., 2007, ApJ, 655, 98

\bibitem[{Navarro, Frenk \& White(1997)Navarro, Frenk, \& White}]{Navarro1997}
Navarro J., Frenk C., White S., 1997, ApJ, 490, 493

\bibitem[{{Nelson} {et~al}\mbox{.}(2014){Nelson}, {Lau}, {Nagai}, {Rudd}, \&
  {Yu}}]{2014ApJ...782..107N}
{Nelson} K., {Lau} E.~T., {Nagai} D., {Rudd} D.~H., {Yu} L., 2014, ApJ, 782,
  107

\bibitem[{{Noh} \& {Cohn}(2012)}]{2012MNRAS.426.1829N}
{Noh} Y., {Cohn} J.~D., 2012, MNRAS, 426, 1829

\bibitem[{{Planck Collaboration} {et~al}\mbox{.}(2014{\natexlab{a}}){Planck
  Collaboration}, {Ade}, {Aghanim}, {Armitage-Caplan}, {Arnaud}, {Ashdown},
  {Atrio-Barandela}, {Aumont}, {Aussel}, {Baccigalupi}, \&
  et~al.}]{2014A&A...571A..29P}
{Planck Collaboration} {et~al.}, 2014{\natexlab{a}}, A\&A, 571, A29

\bibitem[{{Planck Collaboration} {et~al}\mbox{.}(2014{\natexlab{b}}){Planck
  Collaboration}, {Ade}, {Aghanim}, {Armitage-Caplan}, {Arnaud}, {Ashdown},
  {Atrio-Barandela}, {Aumont}, {Baccigalupi}, {Banday}, \&
  et~al.}]{2014A&A...571A..20P}
{Planck Collaboration} {et~al.}, 2014{\natexlab{b}}, A\&A, 571, A20

\bibitem[{{Planck Collaboration} {et~al}\mbox{.}(2015{\natexlab{a}}){Planck
  Collaboration}, {Ade}, {Aghanim}, {Arnaud}, {Ashdown}, {Aumont},
  {Baccigalupi}, {Banday}, {Barreiro}, {Barrena}, \&
  et~al.}]{2015arXiv150201598P}
{Planck Collaboration} {et~al.}, 2015{\natexlab{a}}, ArXiv e-prints

\bibitem[{{Planck Collaboration} {et~al}\mbox{.}(2015{\natexlab{b}}){Planck
  Collaboration}, {Ade}, {Aghanim}, {Arnaud}, {Ashdown}, {Aumont},
  {Baccigalupi}, {Banday}, {Barreiro}, {Bartlett}, \&
  et~al.}]{2015arXiv150201597P}
{Planck Collaboration} {et~al.}, 2015{\natexlab{b}}, ArXiv e-prints

\bibitem[{{Press} {et~al}\mbox{.}(1992){Press}, {Teukolsky}, {Vetterling}, \&
  {Flannery}}]{1992nrfa.book.....P}
{Press} W.~H., {Teukolsky} S.~A., {Vetterling} W.~T., {Flannery} B.~P., 1992,
  {Numerical Recipes in FORTRAN. The Art of Scientific Computing}. Cambridge:
  University Press, |c1992, 2nd ed.

\bibitem[{{Rasia} {et~al}\mbox{.}(2006){Rasia}, {Ettori}, {Moscardini},
  {Mazzotta}, {Borgani}, {Dolag}, {Tormen}, {Cheng}, \&
  {Diaferio}}]{2006MNRAS.369.2013R}
{Rasia} E. {et~al.}, 2006, MNRAS, 369, 2013

\bibitem[{{Rozo} {et~al}\mbox{.}(2014){Rozo}, {Evrard}, {Rykoff}, \&
  {Bartlett}}]{2014MNRAS.438...62R}
{Rozo} E., {Evrard} A.~E., {Rykoff} E.~S., {Bartlett} J.~G., 2014, MNRAS, 438,
  62

\bibitem[{{Rozo} {et~al}\mbox{.}(2009){Rozo}, {Rykoff}, {Evrard}, {Becker},
  {McKay}, {Wechsler}, {Koester}, {Hao}, {Hansen}, {Sheldon}, {Johnston},
  {Annis}, \& {Frieman}}]{2009ApJ...699..768R}
{Rozo} E. {et~al.}, 2009, ApJ, 699, 768

\bibitem[{{Rudd}, {Zentner} \& {Kravtsov}(2008){Rudd}, {Zentner}, \&
  {Kravtsov}}]{2008ApJ...672...19R}
{Rudd} D.~H., {Zentner} A.~R., {Kravtsov} A.~V., 2008, ApJ, 672, 19

\bibitem[{{Sembolini} {et~al}\mbox{.}(2013){Sembolini}, {Yepes}, {De Petris},
  {Gottl{\"o}ber}, {Lamagna}, \& {Comis}}]{2013MNRAS.429..323S}
{Sembolini} F., {Yepes} G., {De Petris} M., {Gottl{\"o}ber} S., {Lamagna} L.,
  {Comis} B., 2013, MNRAS, 429, 323

\bibitem[{{Sievers} {et~al}\mbox{.}(2013){Sievers}, {Hlozek}, {Nolta},
  {Acquaviva}, {Addison}, {Ade}, {Aguirre}, {Amiri}, {Appel}, {Barrientos},
  {Battistelli}, {Battaglia}, {Bond}, {Brown}, {Burger}, {Calabrese},
  {Chervenak}, {Crichton}, {Das}, {Devlin}, {Dicker}, {Bertrand Doriese},
  {Dunkley}, {D{\"u}nner}, {Essinger-Hileman}, {Faber}, {Fisher}, {Fowler},
  {Gallardo}, {Gordon}, {Gralla}, {Hajian}, {Halpern}, {Hasselfield},
  {Hern{\'a}ndez-Monteagudo}, {Hill}, {Hilton}, {Hilton}, {Hincks}, {Holtz},
  {Huffenberger}, {Hughes}, {Hughes}, {Infante}, {Irwin}, {Jacobson},
  {Johnstone}, {Baptiste Juin}, {Kaul}, {Klein}, {Kosowsky}, {Lau}, {Limon},
  {Lin}, {Louis}, {Lupton}, {Marriage}, {Marsden}, {Martocci}, {Mauskopf},
  {McLaren}, {Menanteau}, {Moodley}, {Moseley}, {Netterfield}, {Niemack},
  {Page}, {Page}, {Parker}, {Partridge}, {Plimpton}, {Quintana}, {Reese},
  {Reid}, {Rojas}, {Sehgal}, {Sherwin}, {Schmitt}, {Spergel}, {Staggs},
  {Stryzak}, {Swetz}, {Switzer}, {Thornton}, {Trac}, {Tucker}, {Uehara},
  {Visnjic}, {Warne}, {Wilson}, {Wollack}, {Zhao}, \&
  {Zunckel}}]{2013JCAP...10..060S}
{Sievers} J.~L. {et~al.}, 2013, JCAP, 10, 60

\bibitem[{{Sif{\'o}n} {et~al}\mbox{.}(2015){Sif{\'o}n}, {Battaglia},
  {Menanteau}, {Hasselfield}, {Barrientos}, {Bond}, {Crichton}, {Devlin},
  {D{\"u}nner}, {Hilton}, {Hincks}, {Hlozek}, {Huffenberger}, {Hughes},
  {Infante}, {Kosowsky}, {Marsden}, {Marriage}, {Moodley}, {Niemack}, {Page},
  {Spergel}, {Staggs}, {Trac}, \& {Wollack}}]{2015arXiv151200910S}
{Sif{\'o}n} C. {et~al.}, 2015, ArXiv e-prints

\bibitem[{{Smith} {et~al}\mbox{.}(2015){Smith}, {Mazzotta}, {Okabe}, {Ziparo},
  {Mulroy}, {Babul}, {Finoguenov}, {McCarthy}, {Lieu}, {Bahe}, {Bourdin},
  {Evrard}, {Futamase}, {Haines}, {Jauzac}, {Marrone}, {Martino}, {May},
  {Taylor}, \& {Umetsu}}]{2015arXiv151101919S}
{Smith} G.~P. {et~al.}, 2015, ArXiv e-prints

\bibitem[{{Stanek} {et~al}\mbox{.}(2010){Stanek}, {Rasia}, {Evrard}, {Pearce},
  \& {Gazzola}}]{2010ApJ...715.1508S}
{Stanek} R., {Rasia} E., {Evrard} A.~E., {Pearce} F., {Gazzola} L., 2010, ApJ,
  715, 1508

\bibitem[{{Sunyaev} \& {Zeldovich}(1972)}]{1972CoASP...4..173S}
{Sunyaev} R.~A., {Zeldovich} Y.~B., 1972, Comments on Astrophysics and Space
  Physics, 4, 173

\bibitem[{{Tinker} {et~al}\mbox{.}(2008){Tinker}, {Kravtsov}, {Klypin},
  {Abazajian}, {Warren}, {Yepes}, {Gottl{\"o}ber}, \&
  {Holz}}]{2008ApJ...688..709T}
{Tinker} J., {Kravtsov} A.~V., {Klypin} A., {Abazajian} K., {Warren} M.,
  {Yepes} G., {Gottl{\"o}ber} S., {Holz} D.~E., 2008, ApJ, 688, 709

\bibitem[{{von der Linden} {et~al}\mbox{.}(2014){von der Linden}, {Mantz},
  {Allen}, {Applegate}, {Kelly}, {Morris}, {Wright}, {Allen}, {Burchat},
  {Burke}, {Donovan}, \& {Ebeling}}]{2014MNRAS.443.1973V}
{von der Linden} A. {et~al.}, 2014, MNRAS, 443, 1973

\bibitem[{{White}, {Cohn} \& {Smit}(2010){White}, {Cohn}, \&
  {Smit}}]{2010MNRAS.408.1818W}
{White} M., {Cohn} J.~D., {Smit} R., 2010, MNRAS, 408, 1818

\bibitem[{White \& Hu(2000)}]{White2000}
White M., Hu W., 2000, ApJ, 537, 1

\bibitem[{{Wright} \& {Brainerd}(2000)}]{2000ApJ...534...34W}
{Wright} C.~O., {Brainerd} T.~G., 2000, ApJ, 534, 34

\bibitem[{{Yang}, {Bhattacharya} \& {Ricker}(2010){Yang}, {Bhattacharya}, \&
  {Ricker}}]{2010ApJ...725.1124Y}
{Yang} H.-Y.~K., {Bhattacharya} S., {Ricker} P.~M., 2010, ApJ, 725, 1124

\bibitem[{{Yu}, {Nelson} \& {Nagai}(2015){Yu}, {Nelson}, \&
  {Nagai}}]{2015ApJ...807...12Y}
{Yu} L., {Nelson} K., {Nagai} D., 2015, ApJ, 807, 12

\end{thebibliography}

\end{document}